\documentclass{article}

\usepackage{PRIMEarxiv}

\usepackage[utf8]{inputenc} %
\usepackage[T1]{fontenc}    %
\usepackage{hyperref}       %
\usepackage{url}            %
\usepackage{booktabs}       %
\usepackage{amsfonts}       %
\usepackage{nicefrac}       %
\usepackage{microtype}      %
\usepackage{lipsum}
\usepackage{fancyhdr}       %
\usepackage{graphicx}       %
\usepackage{subcaption}
\usepackage{caption}       %
\graphicspath{{figures/}}     %
\usepackage{epstopdf}
\usepackage[table,xcdraw]{xcolor}
\usepackage[table]{xcolor}  %
\usepackage{amsmath}  %

\usepackage{multirow}
\usepackage{placeins}
\usepackage{soul,color}
\title{Improving realistic material property prediction using domain adaptation based machine learning
\thanks{\textit{\underline{Citation}}: 
\textbf{Jeffrey Hu et al.. MaterialDA. 15 Pages.... DOI:000000/11111.}} 
}

\author{
Jeffrey Hu\\
Dutch Fork High School, Irmo, SC, 39063\\
 Department of Computer Science and Engineering\\
  University of South Carolina\\
  Columbia, SC 29201   
   \And
David Liu\\
 Department of Electrical Engineering and Computer Science
 \\
  University of Michigan\\
  Ann Arbor, MI 48103 \\  
  \And
 Nihang Fu\\
 Department of Computer Science and Engineering\\
  University of South Carolina\\
  Columbia, SC 29201 \\  
  \And
 Rongzhi Dong\\
 Department of Computer Science and Engineering\\
  University of South Carolina\\
  Columbia, SC 29201 \\  
}

\begin{document}
\maketitle

\begin{abstract}

Materials property prediction models are usually evaluated using random splitting of datasets into training and test datasets, which not only leads to over-estimated performance due to inherent redundancy, typically existent in material datasets, but also deviate away from the common practice of materials scientists: they are usually interested in predicting properties for a known subset of related out-of-distribution (OOD) materials rather than a universally distributed samples. Feeding such target material formulas/structures to the machine learning models should improve the prediction performance while most current machine learning (ML) models neglect this information. Here we propose to use domain adaptation (DA) to enhance current ML models for property prediction and evaluate their performance improvements in a set of five realistic application scenarios. Our systematic benchmark studies show that there exist DA models that can significantly improve the OOD test set prediction performance while standard ML models and most of the other DAs cannot improve or even deteriorate the performance. Our benchmark datasets and DA code can be freely accessed at \url{https://github.com/Little-Cheryl/MatDA}.

\end{abstract}

\keywords{material property prediction \and out of distribution \and domain adaptation \and domain shift \and machine learning}

\section{Introduction}

Nowadays, machine learning (ML) models are being widely used in materials property prediction for discovering novel materials such as super-hard materials \cite{avery2019predicting,ojih2023screening}, wide band gap materials \cite{xin2021active}, and energy materials \cite{chen2020critical}. A large number of innovations have been proposed to improve the ML performance for materials property prediction, including more expressive descriptors \cite{himanen2020dscribe}, better deep learning models (IRNET)\cite{jha2019irnet}, graph neural networks that better capture interatomic interactions \cite{omee2022scalable,klipfel2023equivariant,kaba2022equivariant,choudhary2021atomistic}, data augmentation \cite{gibson2022data}, multi-fidelity datasets that combine computational and experimental data \cite{chen2021learning}, active learning\cite{rohr2020benchmarking,xin2021active}, and transfer learning\cite{wu2019machine}. These models and algorithms have significantly improved prediction performance over the past few years. However, it has been found that existing ML algorithms have low generalization performance for test samples with different data distributions, and their prediction performance is often over-estimated due to the high dataset redundancy \cite{li2023redundancy} as many materials are accumulated as a result of a tinkering material discovery process over history. Previously the ML-based material property prediction performances are all evaluated by randomly splitting the whole dataset into training and testing sets. The resulting test set does not share a high degree of homogeneity in terms of composition, structure, or properties, but is randomly distributed in the whole dataset space. This practice does not reflect the realistic application scenario for these ML models when they are more likely to be applied to predict the properties for a set of similar materials that have a different distribution from the training set, and are located in the sparse chemical space with few known materials, or tend to have extreme property values. Moreover, current ML and deep learning models do not consider the query material information in the training of the ML models while in practice the compositions or structures of interest are already known and can be incorporated into ML model training to improve the prediction performance. In real cases, researchers are usually interested in a small number of outlier or out-of-distribution (OOD) materials with different data distributions and with similarities in composition, structure, or properties to maximize a specific function. 

There are several related works regard to OOD prediction problems in materials science. Several studies have found that the inherent high redundancy of materials dataset and the random train-test splitting-based evaluation methods have led to over-estimation of ML performance for material property prediction \cite{meredig2018can,xiong2020evaluating}. It has also been found that regular ML models usually have low generalization performance for OOD samples \cite{loftis2020lattice,li2023critical}. For example, Li et al. \cite{li2023critical} found that ML models trained on Materials Project 2018 can have significantly degraded performance on new materials in Materials Project 2021 due to the distribution shift. In the field of machine learning, OOD generalization due to distribution shift between the source domain and target domain has been intensively investigated recently within the context of transfer learning \cite{wenzel2022assaying}, domain generalization \cite{wang2022generalizing,shen2021towards}, causal learning \cite{scholkopf2021toward}, and domain adaptation \cite{wilson2020survey}. In particular, domain adaptation (DA) methods have been widely used in computer vision, medical imaging, and natural language understanding for improving OOD prediction with great success \cite{zhou2022domain,farahani2021brief,yu2023comprehensive}. Based on whether the model is trained with labels of some test set samples, DA methods can be classified into unsupervised DAs (only Xt, the input information of the test samples are used) and supervised DAs (both Xt and a few Yt test samples are used). Based on the working principles, DAs can be divided into feature-based, instance-based, and transfer-learning methods \cite{de2021adapt}.

Despite the obvious advantage to consider target set distribution for ML model training, currently, there are only a few DA applications in science domains, including those in the bioinformatics \cite{abbasi2020deepcda}, health informatics \cite{anastopoulos2021patient}, and chemistry \cite{jin2020adaptive,wu2022metric}. To our best knowledge, there are no such DA applications to solve OOD problems in materials science, e.g. the OOD material property prediction problem, which has its own unique feature characteristics, domain shift relationships, and domain generalization patterns, so that specialized DA methods are needed to further improve the ML performance.

This paper aims to investigate practically realistic ML models for materials property prediction focusing on predicting properties of minority/outlier material clusters that have different distributions with the training set, all of which have the key symptom of OOD machine learning. We then propose and evaluate three categories of domain adaptation methods for solving this problem, including feature-based, instance-based, and parameter-based algorithms. Our extensive benchmark experiments over five OOD test sets categories brought key insights in applying domain adaptation to improved materials property prediction.

Our contributions are summarized as follows:
\begin{itemize}
    \item We proposed a set of benchmark realistic material property prediction problems, which share the characteristic of predicting the property of a set of OOD samples. 
    \item We suggested incorporating the test sample input (composition or structure) into the ML model training process to improve the prediction performance
    \item We applied and evaluated a series of existing domain adaptation models to the composition and structure based materials property prediction and found key insights as regards how to achieve better OOD prediction performance using suitable DAs.
\end{itemize}

\section{Method}
\label{sec:headings}

\subsection{OOD Benchmark problems and datasets}

We downloaded two datasets (matbench\_expt\_gap and matbench\_glass) from the Matbench \cite{dunn2020benchmarking} site including one classification problem for glass materials, and one regression problem related to band gap prediction. 
The raw dataset details are shown in Table \ref{tab:datasets}

\begin{table}[th]
\caption{Raw datasets}
\centering
\label{tab:datasets}
\begin{tabular}{c|c|c|c|c}
\hline
\rowcolor[HTML]{C0C0C0} 
dataset          & task type       & feature                       & \multicolumn{1}{l|}{\cellcolor[HTML]{C0C0C0}sample no}  
 & performance metric \\ \hline
glass            & \cellcolor[HTML]{FFFFFF}classification & composition/Magpie & 5680                                                   & balanced accuracy  \\ \hline
bandgap          & regression & composition/Magpie                            & 4604                                                   & MAE                \\ \hline
\end{tabular}
\end{table}

Instead of evaluating the ML models using standard train-test random splitting or the k-fold cross-validation (also based on random splitting), we propose five realistic material property prediction scenarios, in which the researchers are usually interested in properties of minority or sparse materials. For each raw dataset (band gap or glass), we define five ways to determine which samples from the sparse composition or property space will be selected as the target test samples. 

\subsubsection{Target set generations}

\begin{figure}[ht!] 
    \begin{subfigure}[t]{0.33\textwidth}
        \includegraphics[width=0.99\textwidth]{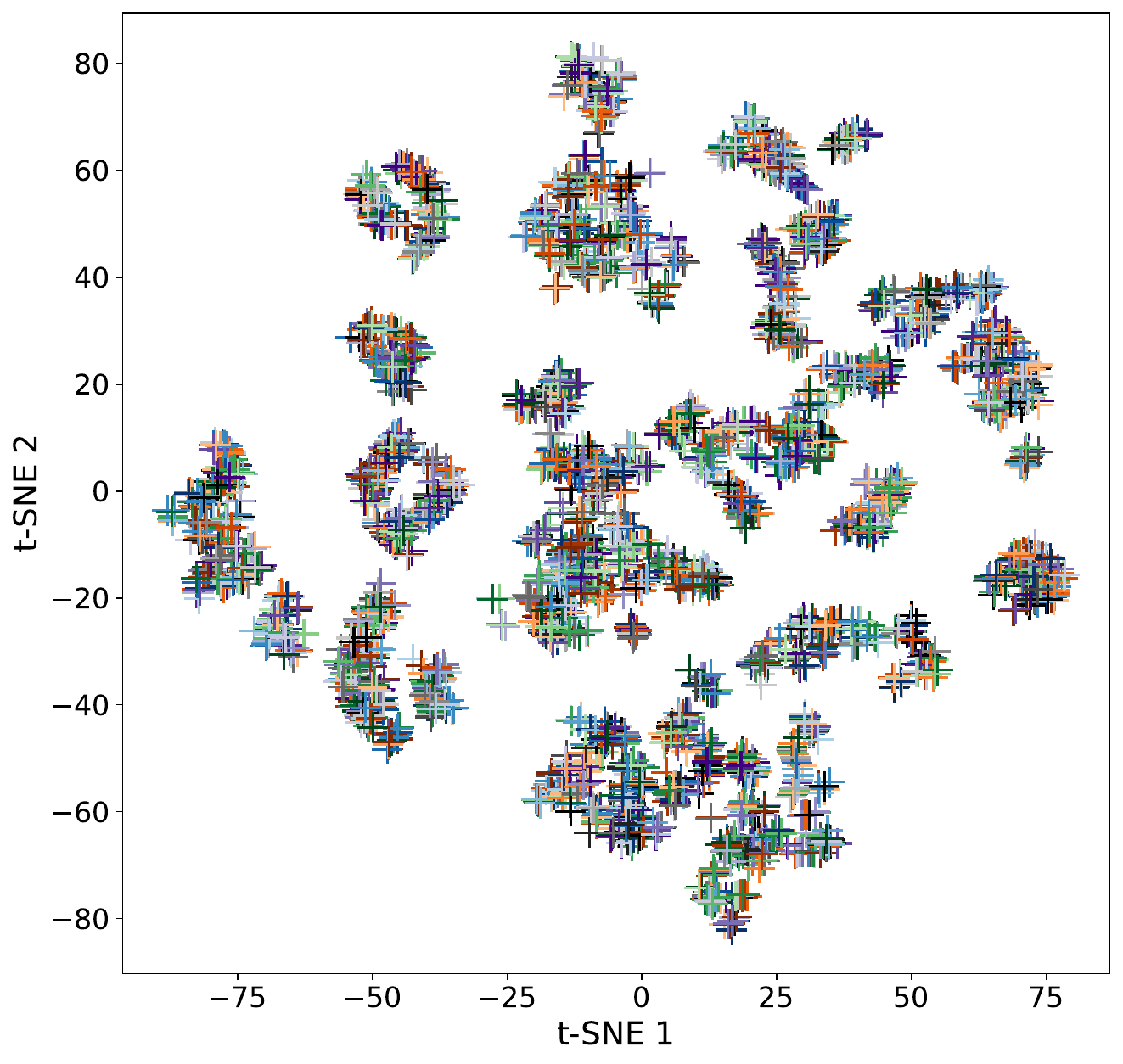}
        \caption{50-fold CV with Random splitting}
        \vspace{3pt}
        \label{fig:whole}
    \end{subfigure}
 \begin{subfigure}[t]{0.33\textwidth}
        \includegraphics[width=0.99\textwidth]{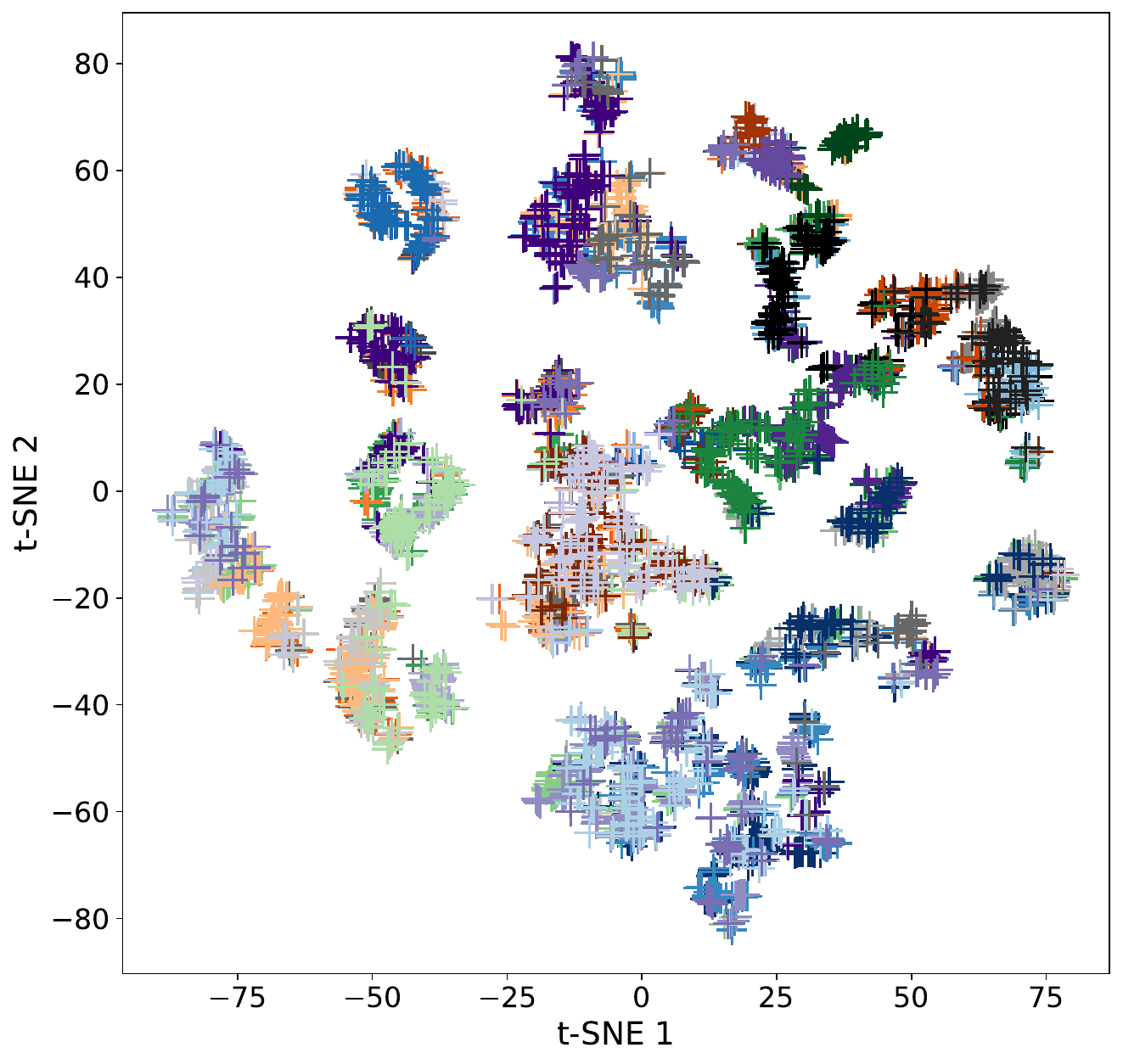}
        \caption{LOCO}
        \vspace{-3pt}
        \label{fig:loco}
    \end{subfigure}    
    \begin{subfigure}[t]{0.33\textwidth}
        \includegraphics[width=0.99\textwidth]{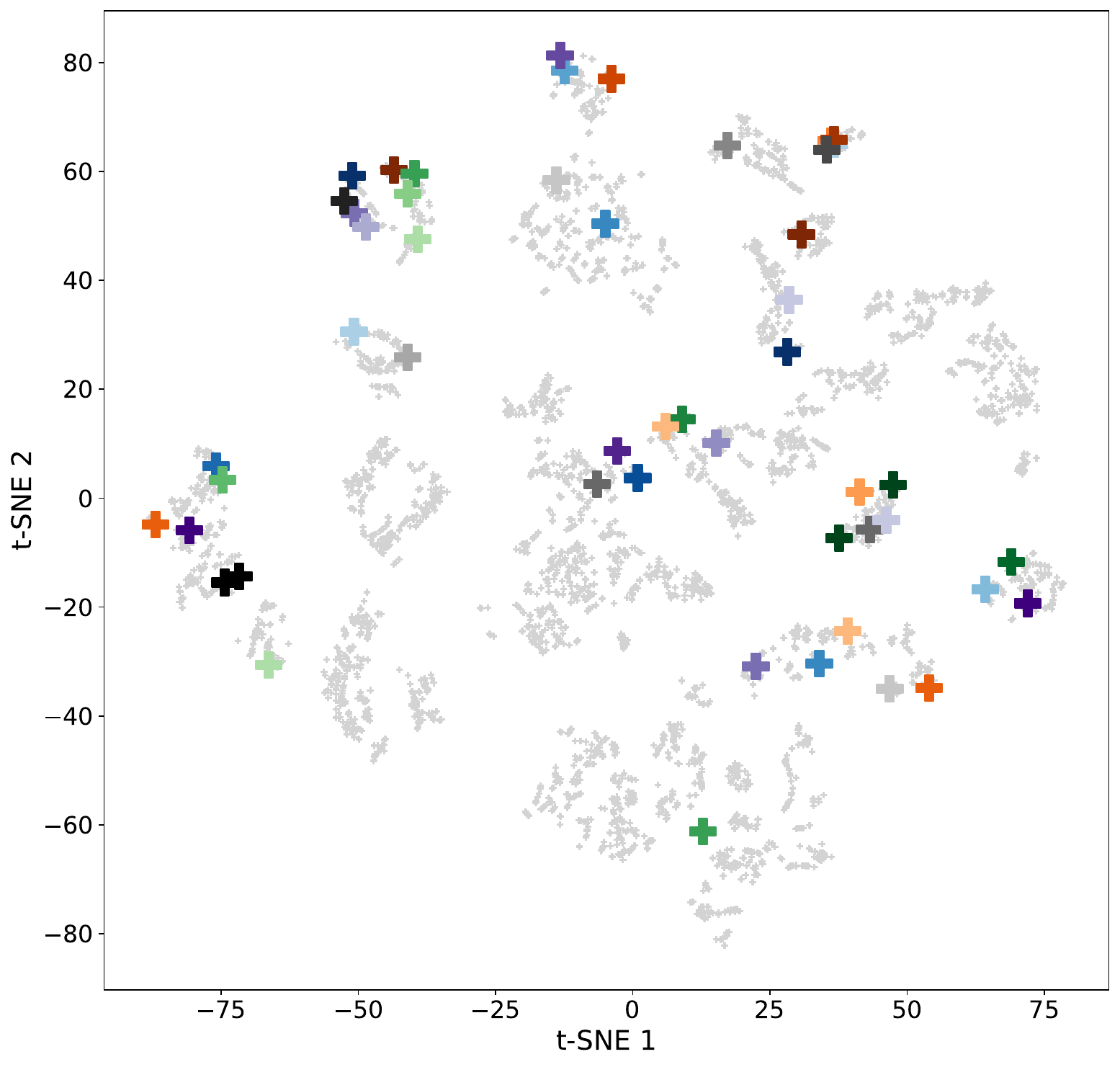}
        \caption{SparseXSingle samples}
        \vspace{3pt}
        \label{fig:sparse_x_single}
    \end{subfigure} 
 \begin{subfigure}[t]{0.33\textwidth}
        \includegraphics[width=0.99\textwidth]{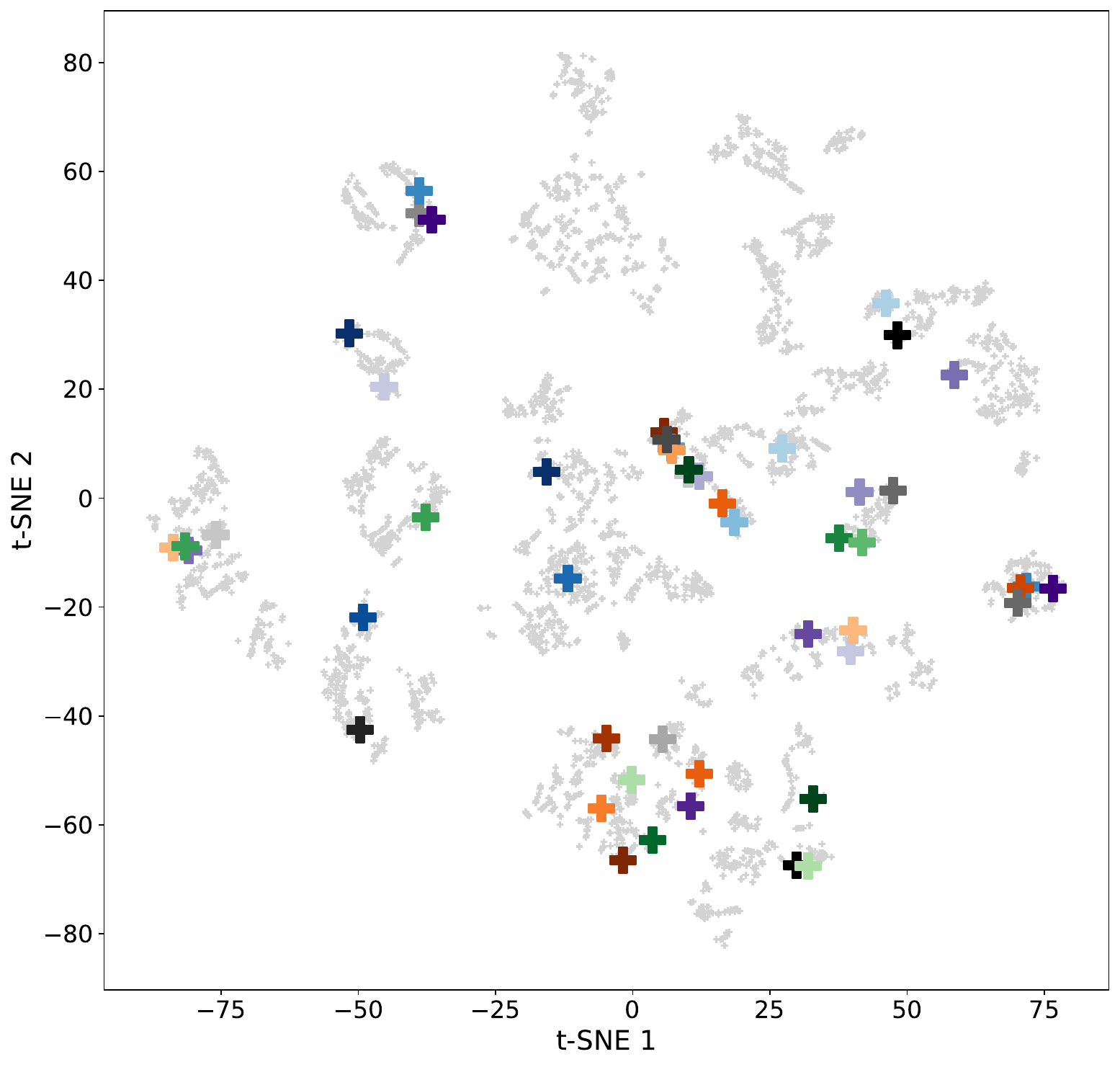}
        \caption{SparseYSingle samples}
        \vspace{3pt}
        \label{fig:sparse_y_single}
    \end{subfigure}              
    \begin{subfigure}[t]{0.33\textwidth}
        \includegraphics[width=0.99\textwidth]{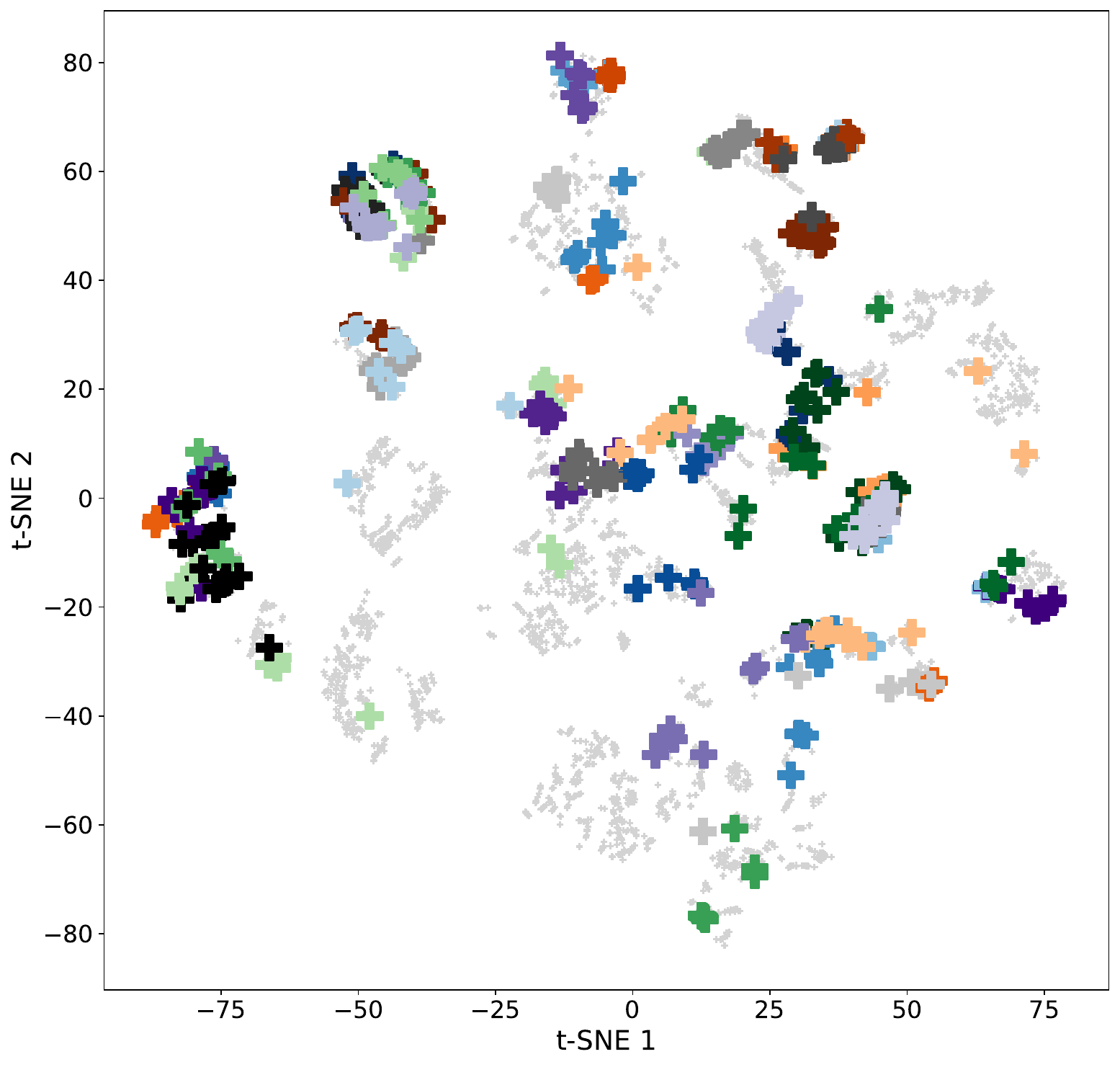}
        \caption{SparseXCluster samples}
        \vspace{-3pt}
        \label{fig:sparse_x_cluster}
    \end{subfigure}
    \begin{subfigure}[t]{0.33\textwidth}
        \includegraphics[width=0.99\textwidth]{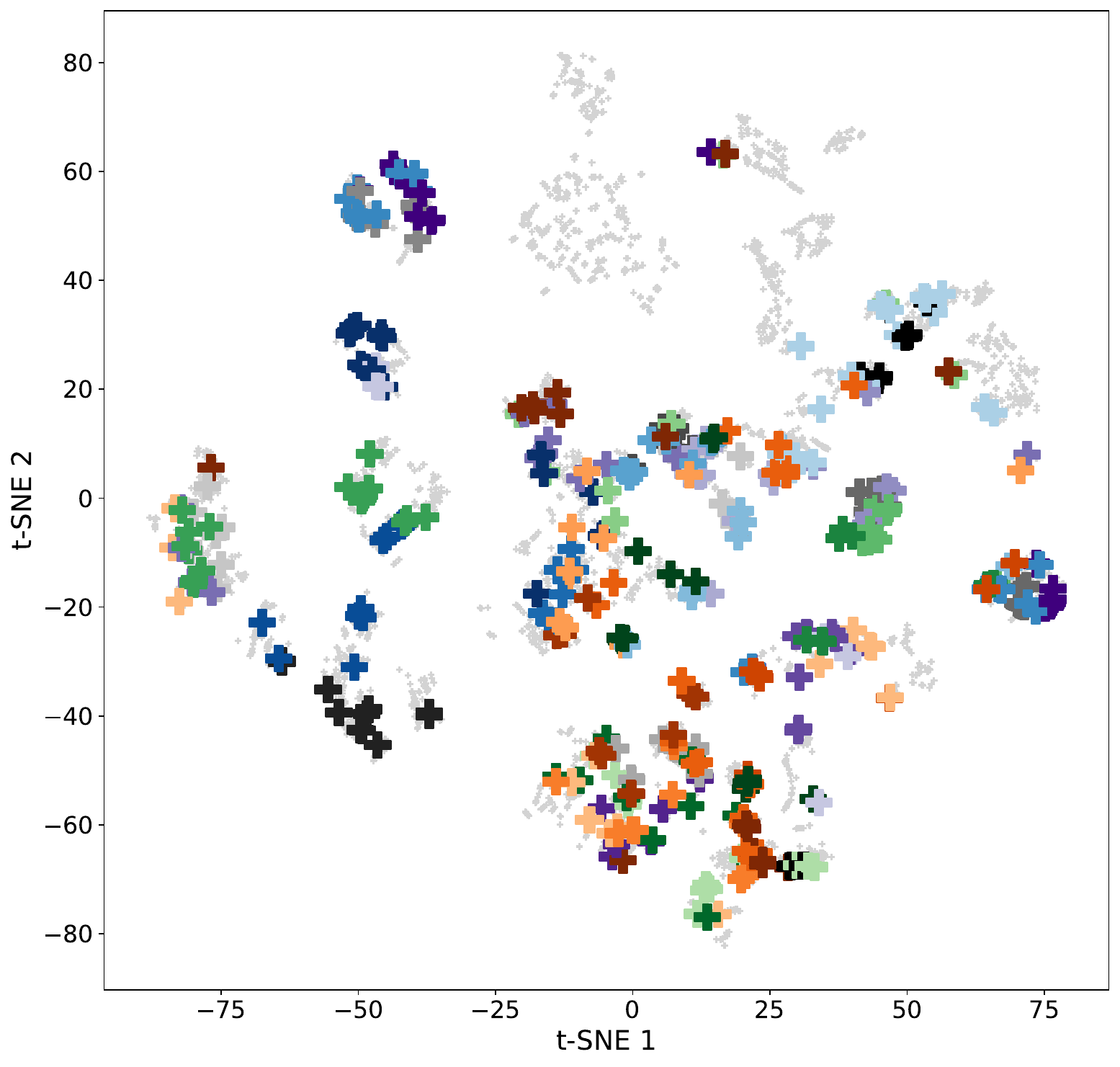}
        \caption{SparseYCluster samples}
        \vspace{-3pt}
        \label{fig:sparse_y_cluster}
    \end{subfigure} 

   \caption{Distribution of standard cross-validation (CV) test set and five OOD test sets using different target generation methods for the band gap dataset. (a) 50-fold CV (with random splitting) of the whole band gap dataset with 4,604 samples represented by cross symbols with 50 different colors. (b) Leave-one-cluster-out target (LOCO) clusters. (c) 50 test samples in SparseXSingle represented by cross symbols with 50 different colors, and grey points represent the remaining samples. (d) 50 test samples in SparseYSingle represented by cross symbols with 50 different colors, and grey points represent the remaining samples. (e) 50 test clusters in SparseXCluster represented by cross symbols with 50 different colors, and grey points represent the remaining samples. (f) 50 test clusters in SparseYCluster represented by cross symbols with 50 different colors, and grey points represent the remaining samples. }
  \label{fig:distribution}
\end{figure}

In most real scenarios, researchers know their target materials of interest and usually have no related labeled samples, which corresponds to the case of unsupervised DA. In this work, we focus on the case that the target set has no labeled samples. We have then proposed the following target set generation methods to emulate the real cases for material property prediction.

\paragraph{Leave-one-cluster-out (LOCO)}
This method was suggested by Meredig et al. \cite{meredig2018can} in their evaluation of the generalization performance of ML models for material property prediction. We first cluster the whole dataset using Magpie features into 50 clusters and then we use each of the clusters as the test sets in turn to evaluate the model performance. Despite it improves the commonly used random splitting method to avoid performance over-estimation, it still counts all samples (including those located in highly dense redundant areas). So it is subject to the over-estimation issue to a certain degree.

\paragraph{Single-point targets with lowest composition density(SparseXSingle)}

In this method, we first convert all dataset compositions into the 132-dimension Magpie feature space and then we apply the t-SNE based dimension reduction to reduce it to 2D space. We then calculate the density for each data point and pick the top 500 least dense samples and we apply K-means clustering to convert it into 50 clusters. Afterward, we then pick one sample out of each cluster, obtaining 50 target samples as our test set. 

\paragraph{Single-point targets with lowest property density(SparseYSingle)}

In this method, we first sort the samples by their y values and calculate the density for each data point's y value and then pick the top 500 least dense samples and we apply K-means clustering to convert it into 50 clusters. We then pick one sample out of each cluster, obtaining 50 target samples as our test set.

\paragraph{Cluster targets with lowest composition density(SparseXCluster)}

This sparse cluster target set generation method is similar to the above-mentioned SparseXSingle method except that after K-means clustering, instead of picking one sample, we further pick N nearest neighbors for each picked sample to form a target cluster. The neighbor-picking process is conducted so that no sample can be selected into multiple target clusters. The neighbors are defined based on the Euclidean distance of Magpie features.

\paragraph{Cluster targets with lowest property density(SparseYCluster)}

This sparse cluster target set generation method is similar to the above-mentioned SparseYSingle method except that after K-means clustering, instead of picking one sample, we further pick N nearest neighbors for each picked sample to form a target cluster. The neighbor-picking process is conducted so that no sample can be selected into multiple target clusters. The neighbors are defined based on the Euclidean distance of Magpie features. In total, we have 50 clusters each with 11 samples in general.

The distribution of the whole band gap datasets and their different target sets are shown in Figure\ref{fig:distribution}. We can find that realistic target sets are more located in sparse areas while commonly used random splitting tend to be located in dense areas with the same distribution as the training set.

In total, we have 10 datasets for DA algorithm evaluations, including bandgap-LOCO, bandgap-SparseXSingle, bandgap-SparseXCluster, bandgap-SparseYSingle, and bandgap-SparseYCluster for regression and glass-LOCO, glass-SparseXSingle, glass-SparseXCluster, glass-SparseYSingle, and glass-SparseYCluster for classification. The number of samples for each cluster of these datasets is shown in Supplementary Table S1.

\subsection{Base algorithms for composition and structure-based materials property prediction}

We use the Random Forest (RF) as the baseline algorithm for domain adaptation evaluation unless specified separately. RF is a strong ML model which can provide a reliable base model for us to build domain adaptation algorithms. For the composition-based property prediction problem (glass classification and band gap prediction in Table\ref{tab:datasets}), we use the Magpie features \cite{ward2016general} as input representation. 
For domain adaptation, we take advantage of the powerful DA package Adapt \cite{de2021adapt}, which has implemented more than 30 DA algorithms. 

For comparison with the DAs with the state-of-the-art algorithms, as reported in the Matbench leaderboard, we evaluate a simple transfer-learning based DA algorithm. For a given target set, we first train a Roost \cite{goodall2020predicting} model using all the training samples. We then select 500 samples similar to the target sample(s) and use them to fine-tune the trained Roost model. The Roost algorithm is an ML approach specifically designed for material property prediction based on material composition. It utilizes a graph neural network framework to learn relationships between material compositions and their corresponding properties. To compare with the performance of traditional strong models, we choose ModNet \cite{de2021materials} algorithm and evaluate its performance on our realistic benchmark datasets and compare with other DA-RF machine learning models.

\subsection{Domain adaptation algorithms}
In many real-world scenarios, machine learning models often suffer from a performance drop when applied to new, unseen data that comes from a distribution different than the training data. This is known as the domain shift problem. Domain adaptation techniques aim to reduce or eliminate this performance degradation by leveraging the knowledge learned from the source and target domain to adapt and generalize well to the target domain. The main idea of domain adaptation is to address the challenge of transferring knowledge learned from a source domain to a target domain when the source and target domains may have different distributions or feature representations. In other words, domain adaptation aims to make a model trained on one dataset (source domain) perform well on another dataset with different characteristics (target domain) without requiring a large amount of labeled data from the target domain by exploiting, e.g., the sample distribution of the target domain.

\begin{figure}[ht!]
  \centering
  \includegraphics[width=0.95\linewidth]{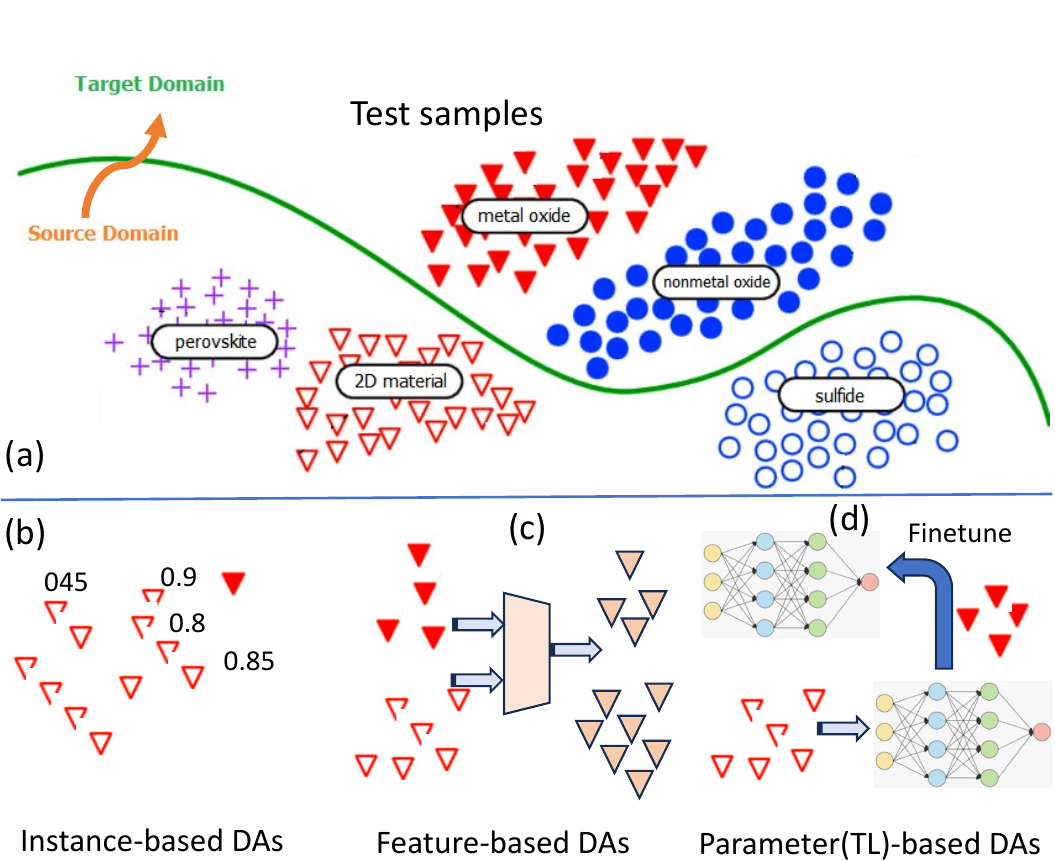}
  \caption{Domain adaptation based machine learning for out-of-distribution material property prediction. (a) OOD prediction problem: model trained with source domain samples is to be used to predict test samples in different target domains; (b) Instance-based DA methods put higher weights on training samples closer to the test samples; (c) feature-based DAs map samples of both source and target domains to a unified representation; (d) parameter or transfer-learning based DAs fine-tune a pre-trained model with a small number of labeled target domain samples. }
  \label{fig:da_framework}
\end{figure}

There are three major categories of domain adaptation algorithms including feature-based, instance-based, and parameter-based methods. Figure \ref{fig:da_framework} (a) shows the out-of-domain prediction problem and the key ideas of three main categories of DA methods. 

\paragraph{Feature-based DA methods}

The feature-based domain adaptation algorithm operates based on the research of common features of a source and target domain. This machine learning technique aims to learn a feature representation that is domain-invariant or transferable. An encoded feature space is created to correct the distributions between the source and target domains. The task on the source and target domain is then learned (aligning feature distributions) in the encoded feature space which permits the generalization of target domains even with limited labeled data. Due to the feature-based domain adaptation's ability to process data quickly and efficiently, this technique encompasses a number of advantages including the ability to utilize labeled source data (even when data is scarce), reduced annotation effort, flexibility (able to align feature representations of source and target domains with different data distributions), improved generalization to a target domain, and overall improved model performance.

We have tested the following feature-based DA methods for the problem: 
\begin{itemize}
    \item Frustratingly easy domain adaptation (PRED) \cite{daume2009frustratingly}
    \item Feature Augmentation (FA) \cite{daume2007frustratingly}
    \item Correlation Alignment (CORAL) \cite{sun2016return}
    \item Subspace Alignment (SA) \cite{fernando2013unsupervised}
    \item Transfer Component Analysis (TCA) \cite{pan2010domain}
    \item Feature Selection with MMD (FMMD) \cite{uguroglu2011feature}
    \item Deep Correlation Assignment (DeepCORAL) \cite{sun2016deep}. We use a default neural network with one hidden layer with 10 neurons and a ReLU activation function. 
\end{itemize}
Out of the seven feature-based DA methods, only the PRED and FA algorithms are supervised DA methods. 
For band gap regression, we used all the above DA methods except FMMD, which took too long to run for this dataset. For the glass classification problem, we used all methods except for DeepCORAL. 

\paragraph{Instance-based DA methods:}
While feature-based domain adaptation aligns with feature representations, instance-based domain adaptation focuses on the transfer of labeled instances from the source domain to the target domain. The core principle of instance-based domain adaptation is to adjust the weight (by multiplying the losses of individual training instances by a positive weight) of labeled training data to correct the differences between source and target distributions. The weight-adjusted training instances are then directly used to learn the task.

We have applied the following instance-based DA algorithms in this study:
\begin{itemize}
    \item Weighting Adversarial Neural Network (WANN) \cite{de2021adversarial}
    \item Kernel Mean Matching (KMM) \cite{huang2006correcting}
    \item Relative Unconstrained Least-Squares Importance Fitting (RULSIF) \cite{yamada2013relative}
    \item Unconstrained Least-Squares Importance Fitting (ULSIF) \cite{kanamori2009least}
    \item Nearest Neighbors Weighting (NNW) \cite{loog2012nearest}
    \item BalancedWeighting (BW) \cite{wu2004improving} 

\end{itemize}

Out of the seven instance-based DA methods, BW and WANN are supervised algorithms while others are unsupervised ones. For the band gap regression problem, we use all the above except ULSIF. For the glass classification problem, we have evaluated all the above DA models.

\paragraph{Parameter-based DA methods:}

In parameter-based DA methods, the parameters of one or a few pre-trained models trained with the source data are adapted to build a fine-tuned model for the task on the target domain. This is a typical scenario of transfer learning. 
We have evaluated the following parameter-based DA algorithms for the band gap prediction and glass classification problems.

\begin{itemize}
    \item Regular Transfer with Linear Regression (RegularTransferLR) \cite{chelba2006adaptation}
    \item Regular Transfer with Neural Network (RegularTransferNN) \cite{chelba2006adaptation}
    \item Linear Interpolation between SrcOnly and TgtOnly (LinInt) \cite{daume2007frustratingly}
    \item TransferTreeClassifer \cite{segev2016learn}
    \item Regular Transfer for Linear Classification (RegularTransferLC) \cite{chelba2006adaptation}
    \item Transfer AdaBoost for Regression (TrAdaBoostR2) \cite{pardoe2010boosting}
    \item Transfer AdaBoost for Classification (TrAdaBoost) \cite{dai2007boosting}
\end{itemize}
All these parameter-based methods are supervised DA methods that need target annotated samples to fine-tune the model. For the band gap regression problem, we evaluated all the above methods except TrAdaBoost, TransferTreeClassifer, and RegularTransferLC.
For the glass classification problem, we evaluated  TransferTreeClassifier, RegularTransferLC, RegularTransferNN, TrAdaBoost, and LinInt.

\subsection{Evaluation criteria}

We use the following performance metrics for evaluating dataset redundancy's impact on model performance, including Mean Absolute Error (MAE), and Balanced Mean Accuracy:

\begin{equation}
\text{MAE} = \frac{1}{n} \sum_{i=1}^{n} \left| y_i - \hat{y}_i \right|    
\end{equation}

\begin{equation}
\text{Balanced Mean Accuracy} = \frac{1}{2} \left(\frac{\text{True Positive Rate}}{\text{Positive Rate}} + \frac{\text{True Negative Rate}}{\text{Negative Rate}}\right)
\end{equation}

Where \(y_i\) represents the observed or true values, \(\hat{y}_i\) represents the predicted values, and \(\bar{y}\) represents the mean of the observed values. The summation symbol \(\sum\) is used to calculate the sum of values, and \(n\) represents the number of data points in the dataset.

\section{Results and Discussion}
\label{sec:others}

\FloatBarrier

\subsection{DA for Leave-one-cluster-out (LOCO) targets}

Compared to the random train-test splitting or the cross-validation, the LOCO targets tend to have different distributions compared to the training sets (See Figure\ref{fig:distribution}f), which leads to increased challenges for the regular machine learning models. Here we evaluate whether different types of domain adaptation methods can be used to improve the generalization performance of the baseline models in the case of a distribution shift. Without special notation, we use the Random Forest algorithm as the baseline model upon which the DA models are applied to. 

\begin{figure*}[ht] 
    \centering
    \begin{minipage}[c]{0.46\textwidth}
        \centering
        \includegraphics[width=\textwidth]{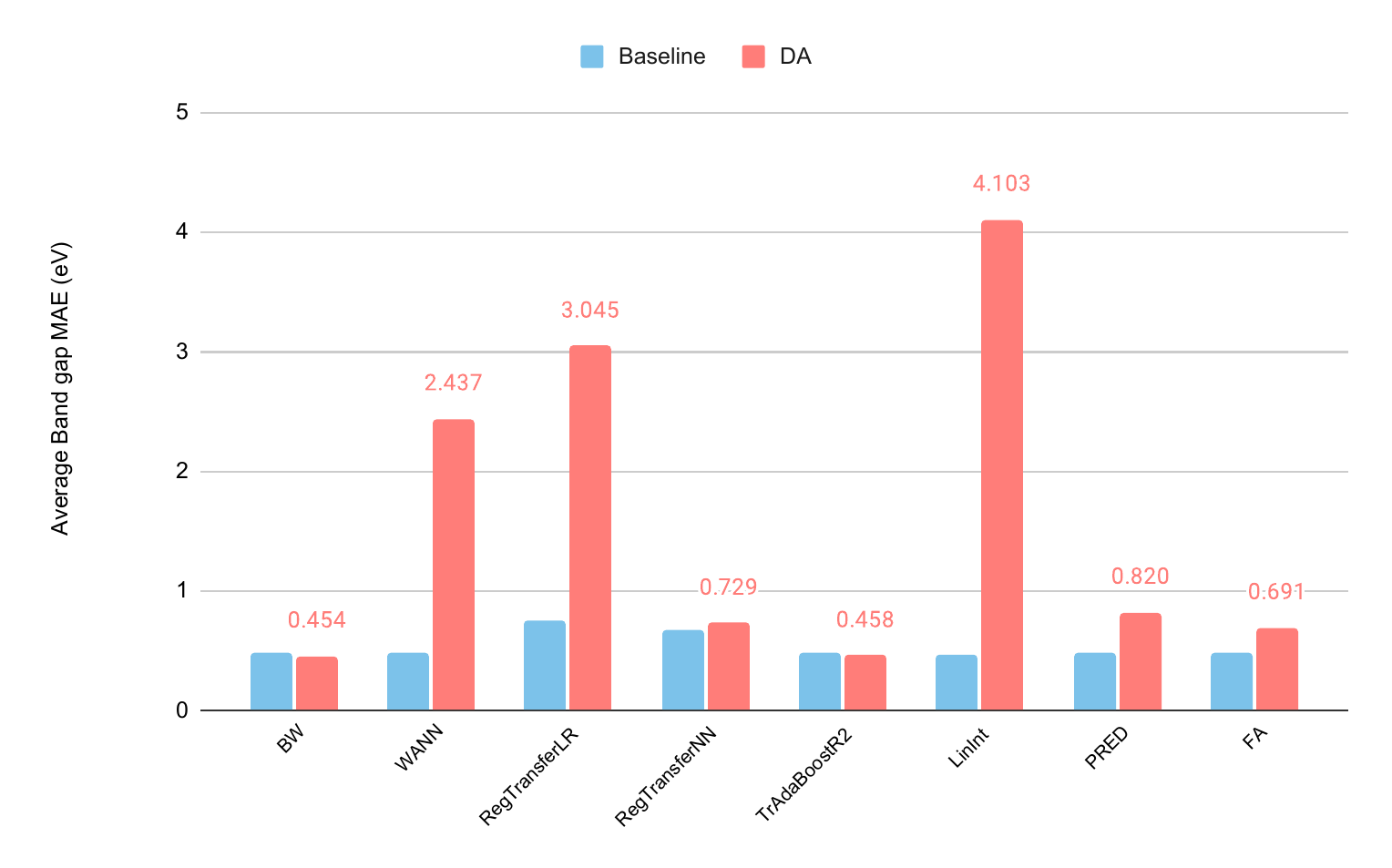}
        \subcaption{Supervised DA methods}
    \end{minipage}
    \begin{minipage}[c]{0.50\textwidth}
        \centering
        \includegraphics[width=\textwidth]{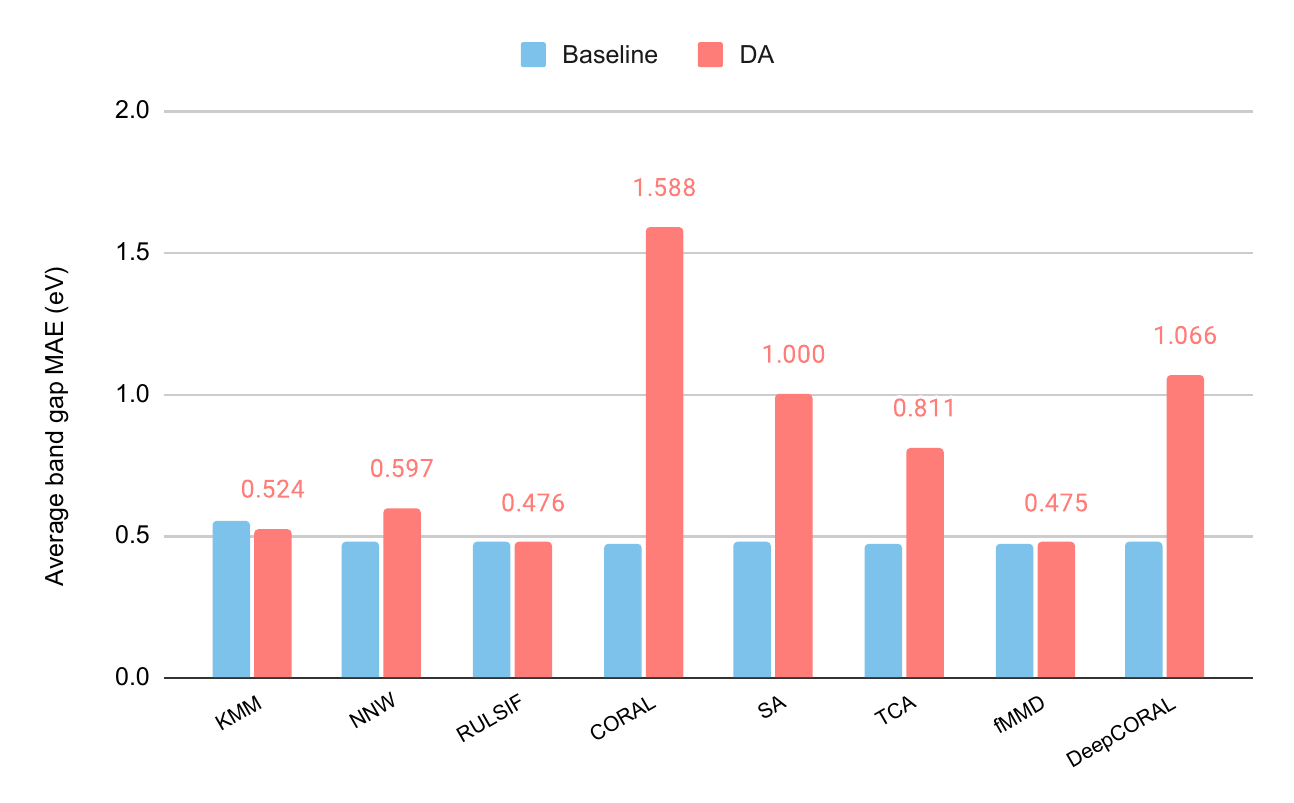}
        \subcaption{Unsupervised DA methods}
    \end{minipage}\\
    
    \caption{Performance of supervised DA models and unsupervised DA models compared with the baseline ML models for band gap prediction over the bandgap-LOCO dataset.  (a) supervised DA versus Random Forest. Only three labeled target samples are used for DA. (b) Unsupervised DAs versus Random Forest.}
    \label{fig:LOCO_bandgap}
    
\end{figure*}

Figure\ref{fig:LOCO_bandgap} (a) shows the results of the supervised DA methods for band gap prediction. We find that the MAE of the supervised DA methods generally increases, delineating that the models' performance becomes worse. Out of 8 supervised DA methods, only the BW and TrAdaBoostR2 have slightly improved with BW's MAE reduced from 0.477 to 0.454 eV and TrAdaBoostR2's MAE reduced from 0.477 to 0.458 eV. In contrast, the methods WANN, RegularTransferLR, and LinInt show significant decreases in performance compared to the baselines. We find that all feature-based DA methods including LinInt, PRED, and FA experienced some decrease in performance. 

Figure\ref{fig:LOCO_bandgap} (b) reflects similar results shown in Figure\ref{fig:LOCO_bandgap} (a) with only two unsupervised DA methods with slightly improved performance, including KMM (MAE reduced from 0.555 eV to 0.524 eV) and RULSIF (MAE reduced from 0.4768 eV to 0.4756 eV). KMM and RULSIF are both instance-based DA models while all the feature-based models decrease in performance (CORAL, SA, TCA, FMMD, DeepCORAL).

Overall, the best performance is achieved by the supervised DA method BW with the lowest MAE of 0.454 eV, a 4.8\% MAE reduction from the baseline. Here we only use three labeled samples in the target cluster for domain adaptation. When we increase this number to 50\% of the test clusters, the BW's MAE can be further reduced to 0.43 eV while TrAdaBoostR2's MAE can be reduced to 0.426 eV. We also compare the RF-based BW performance with those of the two state-of-the-art neural network models for this composition-based band gap prediction problem. First, we find that the ModNet, which is reported to achieve an MAE of 0.3327 eV for the 5-fold cross-validation over the original band gap dataset, only achieves an MAE of 0.8592 eV, a dramatic decrease in its performance, showing its low capability to handle domain shift. In contrast, the Roost \cite{goodall2020predicting} model achieves an MAE of 0.3710 for our Bandgap-LOCO dataset, outperforms ModNet and all other domain adaptation algorithms evaluated so far which are based on random forest or simple neural networks. We further evaluate the unsupervised transfer-learning over the Roost model by fine-tuning the pre-trained model using 500 training samples that are most similar to the test samples. The final MAE is 0.369, a slightly improved result. This demonstrates the importance of the base model for the domain adaptation method.

\begin{figure}[ht]
   \centering
    \begin{minipage}[c]{0.46\textwidth}
        \centering
        \includegraphics[width=\textwidth]{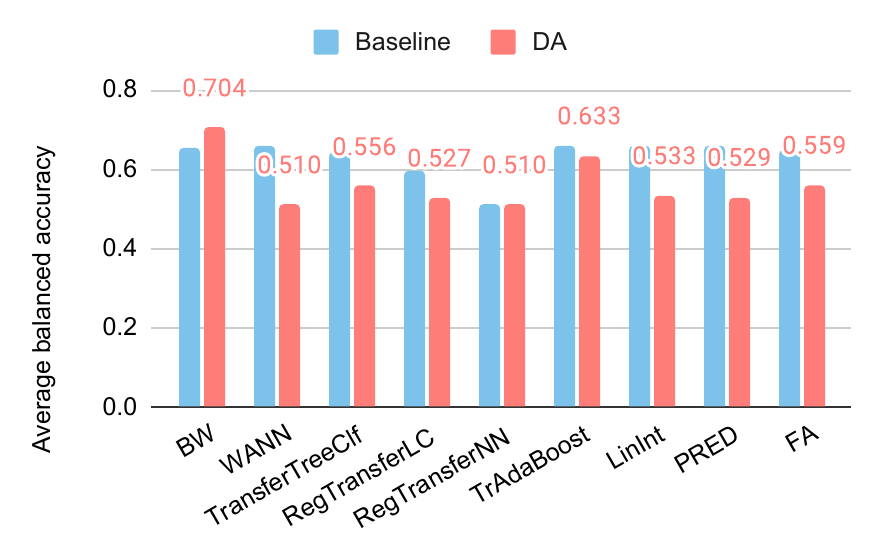}
        \subcaption{Supervised DAs}
    \end{minipage}
    \begin{minipage}[c]{0.46\textwidth}
        \centering
        \includegraphics[width=\textwidth]{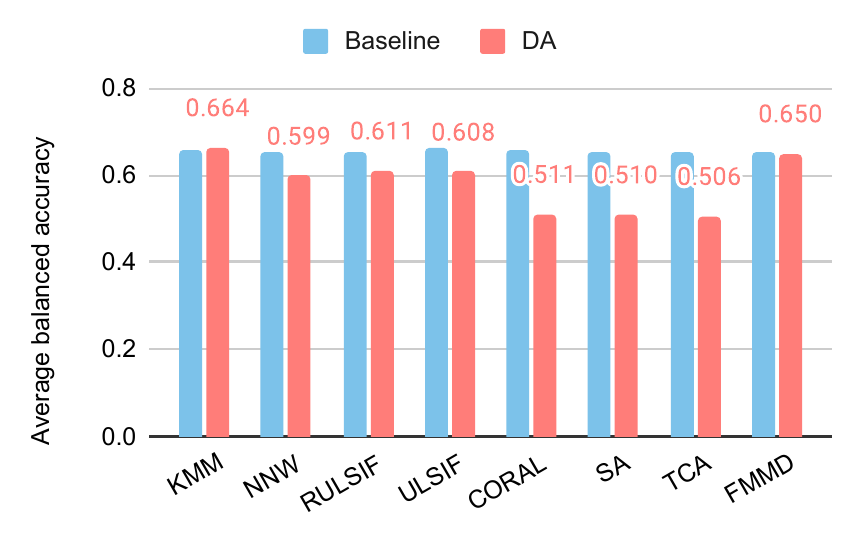}
        \subcaption{Unsupervised DAs}
    \end{minipage}\\
    
  \caption{Performance of DA models for glass classification using DA methods evaluated over the glass-LOCO target clusters. (a) Supervised DA methods (b) Unsupervised DA methods.}
  \label{fig:DA_glass_LOCO}
\end{figure}

Figure\ref{fig:DA_glass_LOCO} (a) shows the results of the supervised DA methods for glass classification. The average balanced accuracy of the supervised DA methods overall decreases, meaning the models' performance was reduced. Of the 9 supervised DA methods, only BW--an instance-based DA method-- improved with an accuracy increasing from 0.652 to 0.704. All other DA methods decrease in performance except for RegularTrasnferNN, whose accuracy remained the same compared to the baseline algorithm. 

Figure\ref{fig:DA_glass_LOCO} (b) shows the average balanced accuracy of unsupervised DA methods (red) versus baseline (blue). Out of the eight unsupervised DA methods, only the KMM shows improvement with its average balanced accuracy increased from 0.652 to 0.704. KMM works by correcting sample bias by minimizing the difference between the means of the source and target domains using the Maximum Mean Discrepancy (MMD) in a reproducing kernel Hilbert space (RKHS). All other unsupervised DA methods decrease in their performance compared to the baseline. We also compared how the number of labeled fine-tuning samples from the target domain affects the DA performance (Supplementary Table S2). We find that when we increase the fine-tuning samples from three to 50\% of the target set, all the DA classification performances improve significantly. This indicates the importance of acquiring more labeled properties for the target domain whenever possible.

\FloatBarrier

\subsection{DA for Single-point targets in sparse X and sparse Y areas}
The single-point sparse X and sparse Y test sets are unique in the sense that there is only one sample in the test set, while all remaining samples are used as training samples. This makes it impossible to apply supervised DA algorithms. Here we evaluate how well the unsupervised DA methods can improve the prediction performance by considering the composition of the single test sample. 

\begin{figure}[ht]
   \centering
    \begin{minipage}[c]{0.46\textwidth}
        \centering
        \includegraphics[width=\textwidth]{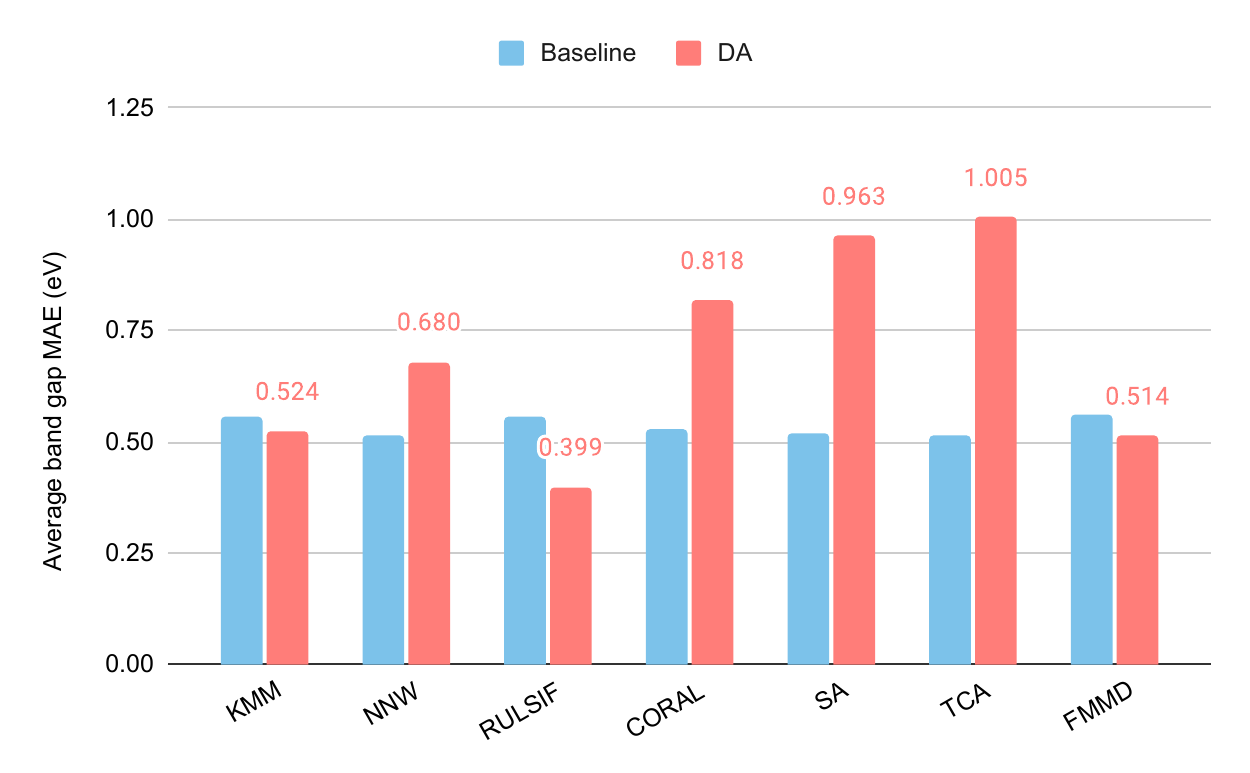}
        \subcaption{Sparse X test sets}
    \end{minipage}
    \begin{minipage}[c]{0.46\textwidth}
        \centering
        \includegraphics[width=\textwidth]{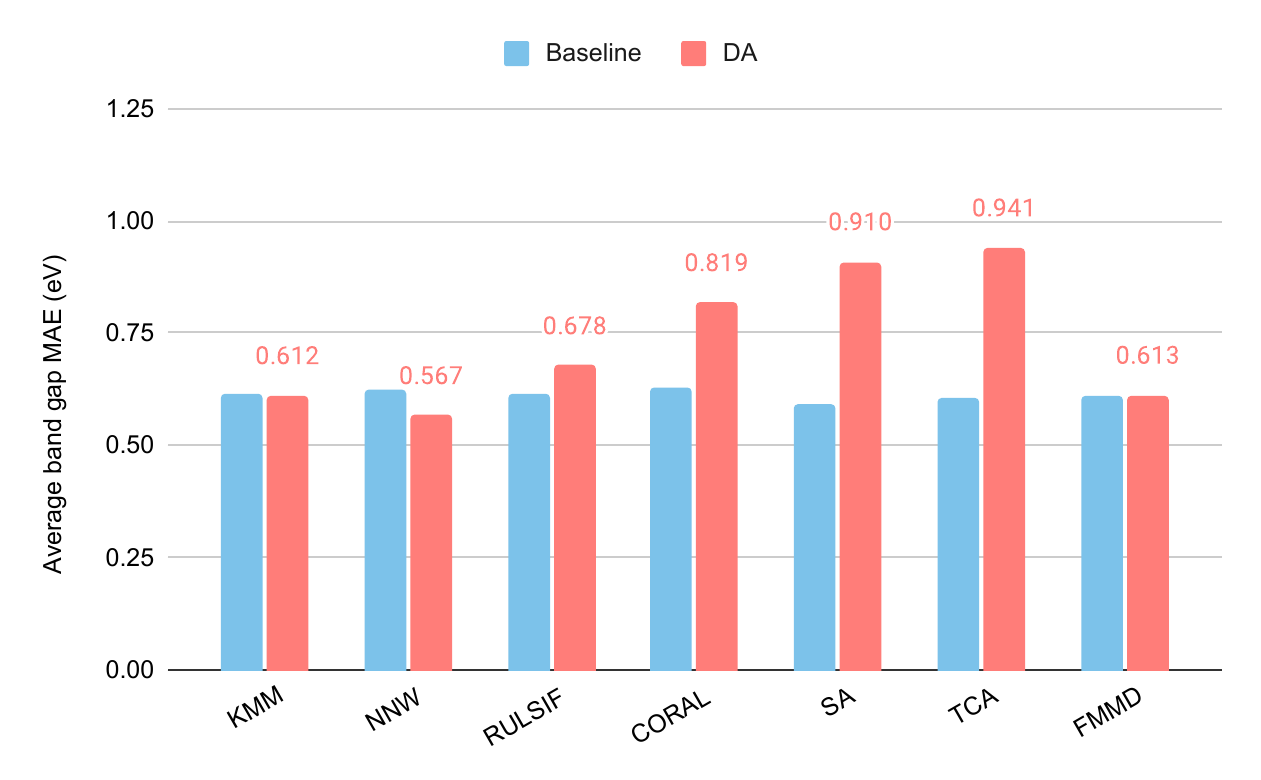}
        \subcaption{Sparse Y test sets}
    \end{minipage}\\
    
  \caption{Performance of unsupervised DA models for band gap prediction evaluated over the 50 single target samples in sparse input (X) and sparse output (Y) areas. (a) Performance of DAs over sparse X test samples (b) Performance of DAs over sparse Y test samples.}
  \label{fig:DA_bandgap_single}
\end{figure}

Figure \ref{fig:DA_bandgap_single} (a) shows the performance of DA methods compared to their baseline for the single sparse X test sets. Out of the seven DA methods, three of them achieve better performance (lower MAEs) including KMM, RULSIF, and FMMD. The largest improvement is by RULSIF which reduces the MAE (0.555 eV) of the baseline RF algorithm to 0.399 eV, a 28\% reduction of the prediction error. In contrast, FMMD and KMM reduce the errors by 8.4\% and 5.7\% respectively. This result demonstrates that the huge potential of DA methods in single-point prediction problems. We also compare the RULSIF's performance over this test set with those of Roost and ModNet and find that it even significantly outperforms Roost with an MAE of 0.483 eV and ModNet with an MAE of 0.436 eV.

We further evaluate the DA performances over the single sparse Y test sets with 50 targets as shown in Figure \ref{fig:DA_bandgap_single} (b). Here only two instance-based DA methods KMM and NNW achieve higher performance than the baseline with an MAE of 0.611 eV and 0.567 eV respectively. All the feature-based DA methods have led to deteriorated performance. Particularly, CORAL, SA, and TCA have significantly higher MAEs compared to the baseline. It is interesting to see that the RULSIF does not work as well as it does for the sparse X test sets.

We then evaluate the DA performances over the Single-X and Single-Y glass datasets (Table \ref{tab:glass_single_results}). First, we find that these two datasets are challenging for the DA methods. For the Single X test sets, only the KMM algorithm has a slight 1.3\% improvement in terms of balanced accuracy. All the other DA methods have either the same or much worse accuracy, especially for the three feature-based DAs such as CORAL, SA, and TCA. For the Singe-Y test sets, the instance-based DA methods NNW and RULSIF improve the baseline performance by 16.7\% and 5.6\% respectively, indicating the potential of DA algorithms for out-of-domain material property prediction.

\begin{table}[tbh]
\centering
\caption{Performance of unsupervised DAs over the glass classification problem. }
\label{tab:glass_single_results}
\begin{tabular}{c|c|c|c|c|c|c|c|c}
\hline
\rowcolor[HTML]{B6D1AD} 
Dataset & Algorithm & \multicolumn{1}{l|}{\cellcolor[HTML]{B6D1AD}KMM} & \multicolumn{1}{l|}{\cellcolor[HTML]{B6D1AD}NNW} & \multicolumn{1}{l|}{\cellcolor[HTML]{B6D1AD}RULSIF} & \multicolumn{1}{l|}{\cellcolor[HTML]{B6D1AD}CORAL} & \multicolumn{1}{l|}{\cellcolor[HTML]{B6D1AD}SA} & \multicolumn{1}{l|}{\cellcolor[HTML]{B6D1AD}TCA} & \multicolumn{1}{l}{\cellcolor[HTML]{B6D1AD}FMMD} \\ \hline
 & Baseline & 0.75 & 0.75 &0.75  & 0.75 & 0.75 & 0.75 & 0.75 \\ \cline{2-9} 
\multirow{-2}{*}{\begin{tabular}[c]{@{}l@{}}Single X\\ test sets\end{tabular}} & DA & \textbf{0.76} & 0.74 & 0.72 & 0.34 & 0.34 & 0.66 & 0.72 \\ \hline
 & Basline & 0.72 & 0.72 & 0.72 & 0.72 & 0.72 & 0.72 & 0.72 \\ \cline{2-9} 
\multirow{-2}{*}{\begin{tabular}[c]{@{}l@{}}Single Y\\ test sets\end{tabular}} & DA & 0.7 & \textbf{0.84} & \textbf{0.76} & 0.44 & 0.44 & 0.58 & 0.7 \\ \hline
\end{tabular}
\end{table}

\FloatBarrier

\subsection{DA for band gap SparseX and SparseY cluster targets}

Test sets of the SparseXCluster and SparseYCluster datasets are constructed by first selecting 50 seed samples with the highest sparsity in the composition (Magpie) or property space, then selecting 10 out of those samples that are most similar to the seed sample. The selected samples are used to evaluate whether an ML model can predict the properties without using closest neighbors. 

\begin{figure}[th] 
    \begin{subfigure}[t]{0.5\textwidth}
        \includegraphics[width=\textwidth]{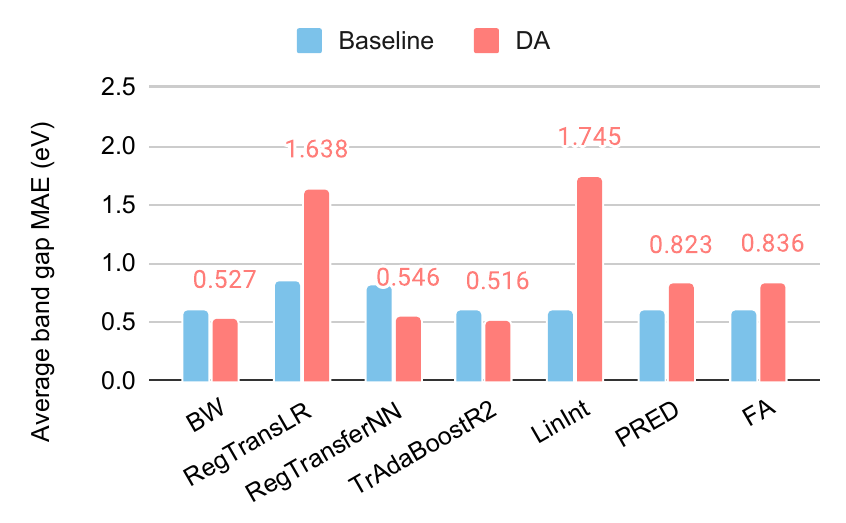}
        \caption{}
        \vspace{-3pt}
        \label{fig:t-Dy4S4Cl4}
    \end{subfigure}\hfill
    \begin{subfigure}[t]{0.5\textwidth}
        \includegraphics[width=\textwidth]{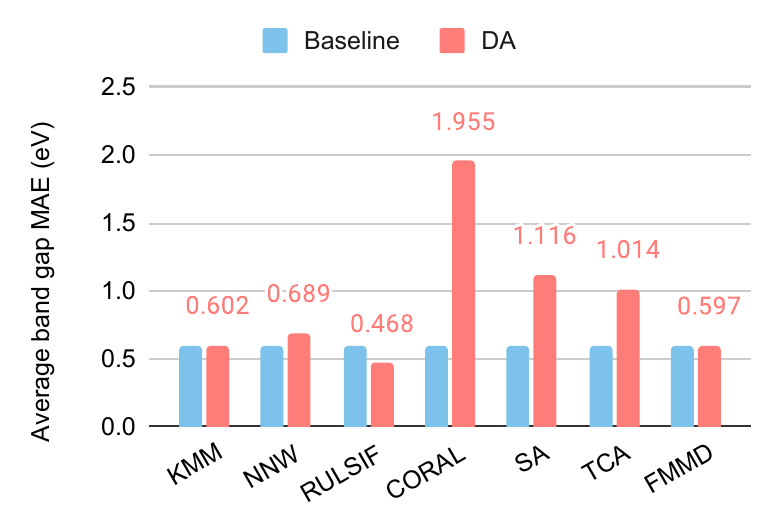}
        \caption{}
        \vspace{-3pt}
        \label{fig:p-Dy4S4Cl4}
    \end{subfigure}
    
    \begin{subfigure}[t]{0.5\textwidth}
        \includegraphics[width=\textwidth]{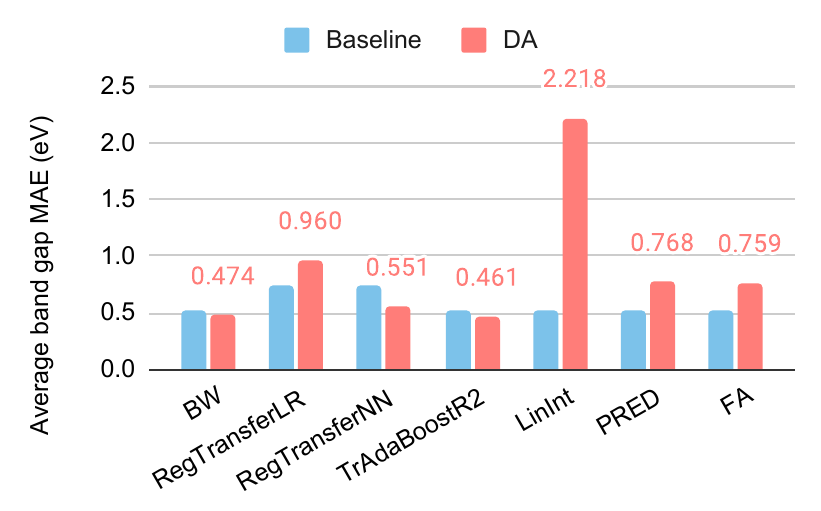}
        \caption{}
        \vspace{-3pt}
        \label{fig:t-As8Ir4}
    \end{subfigure}\hfill
    \begin{subfigure}[t]{0.5\textwidth}
        \includegraphics[width=\textwidth]{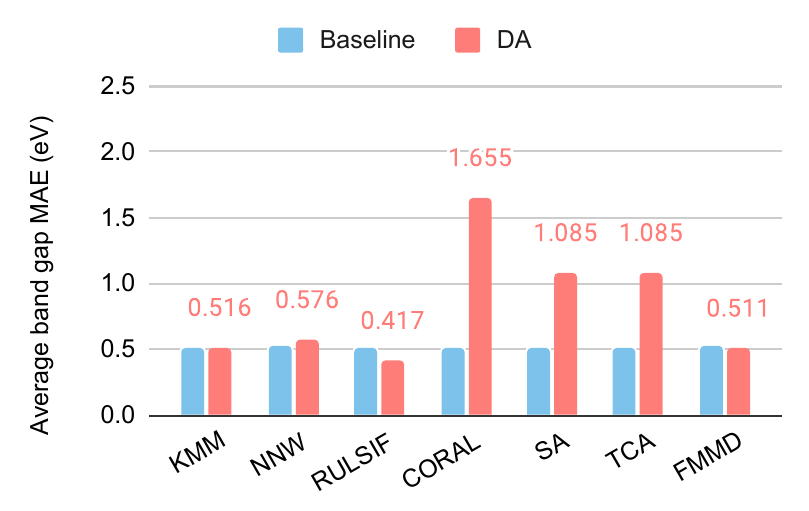}
        \caption{}
        \vspace{-3pt}
        \label{fig:p-As8Ir4}
    \end{subfigure}
    \caption{DA performance for the SparseX and SparseY test clusters derived from the band gap dataset. (a) Supervised DAs for the Sparse X clusters; (b) Unsupervised DAs for the Sparse X clusters; (c) Supervised DAs for the Sparse Y clusters; (b) Unsupervised DAs for the Sparse Y clusters;}
    \label{fig:sparseXY_bandgap}
\end{figure}

Figure \ref{fig:sparseXY_bandgap} (a) and (b) show the performance of supervised and unsupervised DAs over the band gap SparseXCluster dataset. Out of the seven supervised DAs (Figure \ref{fig:sparseXY_bandgap} (a)), three of the DA algorithms have improved performances over their base algorithms. Those DA algorithms are BW, RegularTransferNN, and TrAdaBoostR2 among which the TrAdaBoostR2 achieves the lowest MAE of 0.516 eV. Out of the seven unsupervised DA methods (Figure \ref{fig:sparseXY_bandgap} (b)), only one algorithm-- RULSIF-- has a significant performance improvement over its base RF algorithm with an MAE of 0.468 eV. This is impressive as it beats all the supervised DAs. However, this performance is not as good as the MAE (0.421 eV) of the basic Roost algorithm, showing the power of the neural network model of the Roost.RULSIF's performance, however, is much better than that of the ModNet, which only achieves an MAE of 0.788 eV, a significant degradation from its 0.331 eV for the 5-fold random cross-validation performance as reported in the Matbench.

Figure \ref{fig:sparseXY_bandgap} (c) and (d) show the performance of supervised and unsupervised DAs over the band gap SparseYCluster dataset. Out of the seven supervised DAs (Figure \ref{fig:sparseXY_bandgap} (c)), BW and TrAdaBoostR2 are the only two algorithms to have improved performance over their base algorithms, with TrAdaBoostR2 achieving the lowest MAE of 0.461 eV. Out of the seven unsupervised DA methods (Figure \ref{fig:sparseXY_bandgap} (d)), only one algorithm, RULSIF, has improved performance over the base RF algorithm by 18.7\% with an MAE of 0.417 eV compared to 0.513 eV of the base model. This is impressive as it beats all the supervised DAs and is as good as the MAE (0.419 eV) of the Roost algorithm. RULSIF's performance is also much better than that of the ModNet which only achieves an MAE of 0.824 eV. Overall, we find that the unsupervised RULSIF has demonstrated strong performance for these two challenging test datasets. As an instance-based method for domain adaptation, RULSIF works by correcting the difference between input distributions of source and target domains by finding a source instances reweighting which minimizes the relative Person divergence between source and target distributions.

\FloatBarrier

\subsection{DA for glass sparseXCluster and sparseYCluster datasets}

\begin{figure}[th] 
    \begin{subfigure}[t]{0.5\textwidth}
        \includegraphics[width=\textwidth]{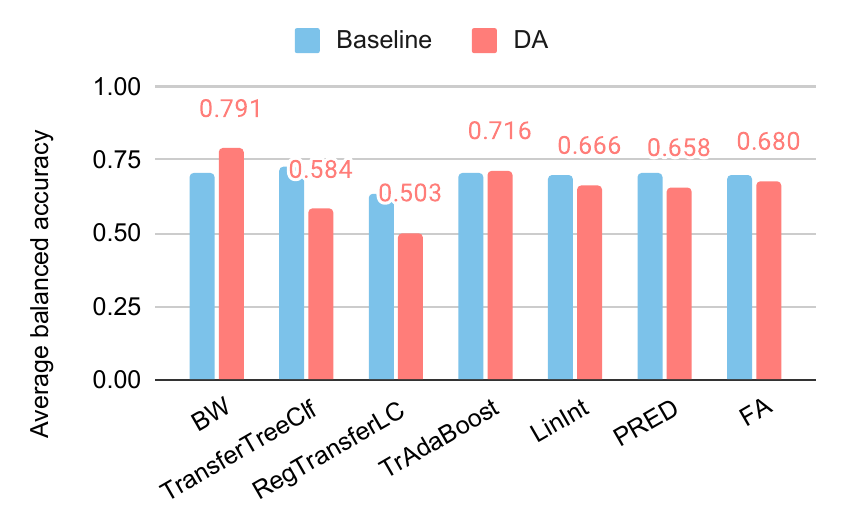}
        \caption{}
        \vspace{-3pt}
        \label{fig:t-Dy4S4Cl4}
    \end{subfigure}\hfill
    \begin{subfigure}[t]{0.5\textwidth}
        \includegraphics[width=\textwidth]{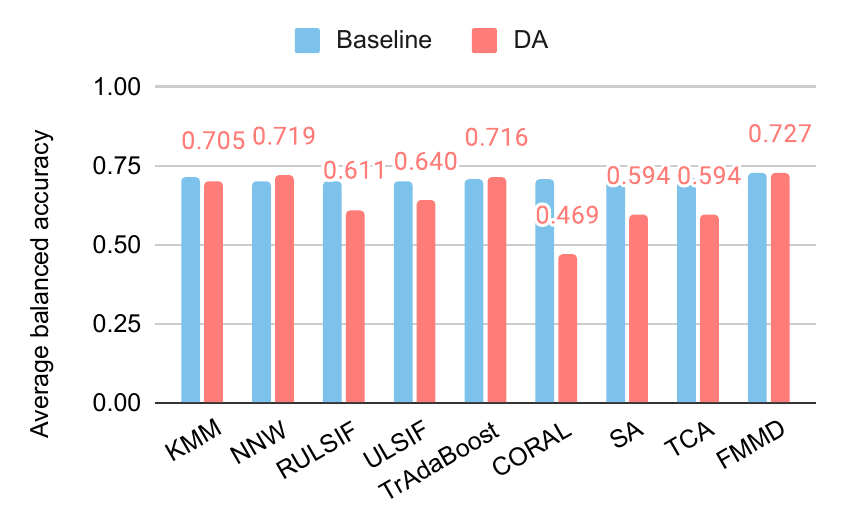}
        \caption{}
        \vspace{-3pt}
        \label{fig:p-Dy4S4Cl4}
    \end{subfigure}
    
    \begin{subfigure}[t]{0.5\textwidth}
        \includegraphics[width=\textwidth]{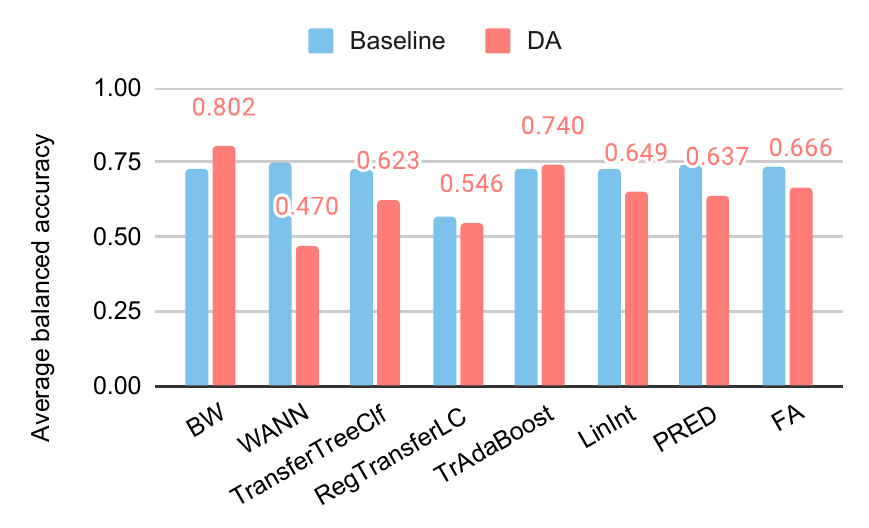}
        \caption{}
        \vspace{-3pt}
        \label{fig:t-As8Ir4}
    \end{subfigure}\hfill
    \begin{subfigure}[t]{0.5\textwidth}
        \includegraphics[width=\textwidth]{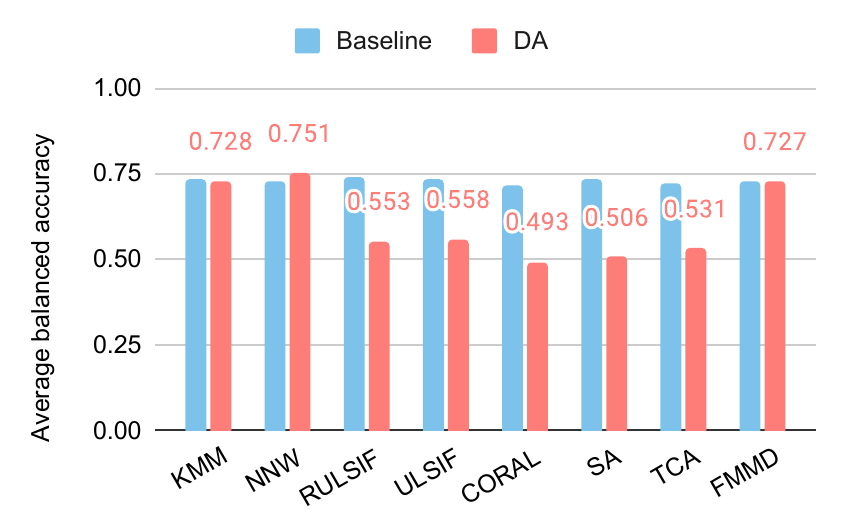}
        \caption{}
        \vspace{-3pt}
        \label{fig:p-As8Ir4}
    \end{subfigure}
    \caption{DA performance for the SparseX and SparseY test clusters derived from the glass dataset. (a) Supervised DAs for the Sparse X clusters; (b) Unsupervised DAs for the Sparse X clusters; (c) Supervised DAs for the Sparse Y clusters; (b) Unsupervised DAs for the Sparse Y clusters;}
    \label{fig:sparseXY_cluster_glass}
\end{figure}

Figure \ref{fig:sparseXY_cluster_glass} (a) and (b) show the performance of supervised DAs and unsupervised DAs over the glass SparseXCluster dataset. Out of the seven supervised DAs (Figure \ref{fig:sparseXY_cluster_glass} (a)), only the BW algorithm has improved accuracy over their base algorithms with a balanced accuracy of 79.1\%. Out of the seven unsupervised DA methods (Figure \ref{fig:sparseXY_cluster_glass} (b)), only one algorithm, NNW, has the performance improvement over its base RF algorithm with an accuracy of 72.9\%. When compared to BW, NNW's performance is significantly lower. We also found that BW has outperformed Roost for this dataset, which has an accuracy of 77.60\%, and ModNet which achieves an accuracy of 60\%, much lower than the 96\% for the same glass dataset but with 5-fold random cross-validation.

Figure \ref{fig:sparseXY_cluster_glass} (c) and (d) shows the performance of supervised DAs and unsupervised DAs over the glass SparseYCluster dataset. Out of the seven supervised DAs (Figure \ref{fig:sparseXY_cluster_glass} (c)), only BW again has improved the performance of their base algorithms with an accuracy of 80.2\%. Out of the seven unsupervised DA methods (Figure \ref{fig:sparseXY_cluster_glass} (d)), only one algorithm, NNW, has improved performance over its base RF algorithm with an accuracy of 75.1\%, which is not as the best supervised DA algorithm BW.
We also found that the BW's performance is better than the Roost algorithm, which achieves an accuracy of 75.5\%. When we further fine-tune the pre-trained Roost model with the 500 most similar training samples to the target set, its performance increases to 77.9\%, which is still below BW's accuracy. This demonstrates the big potential of DA algorithms for these raw sample property predictions. We also ran ModNet over this dataset and found it can only achieve an accuracy of 51\%.  Compared to its 96\% reported on the Matbench, this is a significant degradation, which implies the huge risk to use these models to predict properties for minority or new materials that are located in different areas of the composition or property space.

\FloatBarrier

\subsection{Discussion}

Usually, in realistic materials property prediction, the target compositions or structures are already known, which can be used to guide the ML model training. Moreover, researchers are usually more interested in the properties of novel materials with unusual compositions or properties, leading to a typical out-of-distribution machine learning problem. Here we formulate five OOD material property prediction benchmark datasets for both regression and classification problems and conducted extensive experiments to evaluate how existing well-established domain adaptation methods work in the materials science context. It is found that standard ML models tend to have severely degraded performance for such OOD test sets. While most existing DA models cannot improve the performance of their base model, a few DA algorithms such as BW and RULSIF that capture the true domain shift relationship can achieve much better results compared to the baseline and outperform other state-of-the-art neural network models such as ModNet, demonstrating the huge potential of DA in material property prediction. 

To investigate further DAs for the OOD material property prediction problem, we summarize the best algorithms for each of the 10 datasets evaluated in this study along with their performance scores and standard deviation (Table \ref{tab:best_algo}. First, for the five OOD test sets of band gap prediction, the neural network based Roost algorithm is the best according to its average MAEs over 50 clusters. RULSIF, an instance-based unsupervised DA achieved the best performance for two of the five OOD test sets, which is impressive as its base model is Random Forest rather than neural networks. Another observation is that the standard deviations of all the highest-performing algorithms are relatively high, which could be due to the challenges of this band gap test sets, out of which several clusters are very difficult to predict accurately (Supplementary Figure S1). This is very different from standard K-fold cross-validation experiments or random train-test splitting tests, which tend to have i.i.d test distributions and thus very low-performance variation across different folds. This calls for special attention to practical material property prediction as regards the predicted property values and techniques such as uncertainty quantification \cite{varivoda2023materials} may be introduced to estimate the confidence of the result. 
For the glass datasets, it is found that four DA methods achieve the best classification performance: three of them are done by BW and one by NNW, both of which are instance-based DA methods. 

In this study, the three best-performance DAs are all instance-based methods, indicating their advantages for OOD material property prediction. However, this does not mean that feature-based or transfer-learning based DA models have less potential. More likely, the reason for their low performances is due to their design is currently developed based on assumptions of domain shifts in computer vision, medical imaging, and related field. In contrast, the domain-shift relationship of material properties has unique regularities that the current feature-based DAs cannot model and exploit. The low performance of the transfer-learning (parameter-based) DAs is probably due to the base model we used here are default primitive neural network with just one hidden layer. More powerful network models similar to Roost or ModNet may be used as the pre-trained model to improve their OOD performance. 

While here we only focus on composition-based material property prediction, domain adaptation methods can be easily transferred to structure-based ML Models for materials property prediction. This can be done, e.g., by first training a regular graph neural network model and then using the trained network backbone before the fully connected layer as the feature extractor to convert all structures into latent features for both training and testing samples. We can then apply the DA algorithms to the derived dataset. The pre-trained model based transfer learning can also be used here too.

\begin{table}[tb]
\caption{Best algorithms for each of the ten OOD test datasets}
\label{tab:best_algo}
\begin{tabular}{c|ccc|cc|c}
\hline
\rowcolor[HTML]{B6D1AD} 
 & \multicolumn{3}{c|}{\cellcolor[HTML]{B6D1AD}Bandgap} & \multicolumn{2}{c|}{\cellcolor[HTML]{B6D1AD}Glass} & \multicolumn{1}{c}{\cellcolor[HTML]{B6D1AD}} \\ \hline
\rowcolor[HTML]{B6D1AD} 
dataset & \multicolumn{1}{c|}{\cellcolor[HTML]{B6D1AD}Best algorithm} & \multicolumn{1}{c|}{\cellcolor[HTML]{B6D1AD}MAE (eV)} & \multicolumn{1}{c|}{\cellcolor[HTML]{B6D1AD}std} & \multicolumn{1}{c|}{\cellcolor[HTML]{B6D1AD}Best algorithm} & \multicolumn{1}{c|}{\cellcolor[HTML]{B6D1AD}balanced accuracy} & \multicolumn{1}{c}{\cellcolor[HTML]{B6D1AD}std} \\ \hline
LOCO & \multicolumn{1}{c|}{Roost} & \multicolumn{1}{c|}{0.371} & 0.260 & \multicolumn{1}{c|}{BW} & 0.704 & 0.141 \\ \hline
SparseXCluster & \multicolumn{1}{c|}{Roost} & \multicolumn{1}{c|}{0.421} & 0.414 & \multicolumn{1}{c|}{BW} & 0.791 & 0.209 \\ \hline
SparseYCluster & \multicolumn{1}{c|}{RULSIF} & \multicolumn{1}{c|}{0.417} & 0.314 & \multicolumn{1}{c|}{BW} & 0.802 & 0.187 \\ \hline
{\color[HTML]{222222} SparseXSingle} & \multicolumn{1}{c|}{RULSIF} & \multicolumn{1}{c|}{0.399} & 0.433 & \multicolumn{1}{c|}{Roost} & 0.84 & 0.373 \\ \hline
{\color[HTML]{222222} SparseYSingle} & \multicolumn{1}{c|}{Roost} & \multicolumn{1}{c|}{0.495} & 0.767 & \multicolumn{1}{c|}{NNW} & 0.84 & 0.367 \\ \hline
\end{tabular}
\end{table}

\section{Conclusion}

In real-world materials discovery, researchers already know the target material compositions or structures for which they want to predict their properties. It is desirable to exploit such information to train better ML models for material property prediction in such scenarios. In addition, material scientists usually are more interested in materials with new/rare properties in uncharted design spaces. Here we propose a set of five realistic materials property prediction benchmark problems, in which the test samples are located in sparse composition or property space. We then evaluated the performance of different domain adaptation enhanced machine learning algorithms for the band gap prediction and the glass classification problems. 

Our experiments show that out-of-distribution materials property prediction poses great challenges for regular machine learning algorithms including state-of-the-art algorithms such as ModNet. Out of the three categories of DA methods that we evaluated, the feature-based DAs and the parameter-based DAs (transfer-learning or fine-tuning) do not show an improved performance overall. The reason is two folds: for feature-based DAs, it is probably due to the source-target domain relationships in our materials datasets are different from those in the original DA papers. For the parameter-based DA methods, it may be due to the incompetency of the default base neural network models. Out of all instance-based DA methods, the best DAs are usually supervised DAs. For both categories of DA methods, it seems we have to develop material data-oriented features and transfer-learning DA algorithms that can capture the underlying domain shift relationships. For example, for both the bandgap-LOCO and glass-LOCO test sets, the instance-based supervised BW achieves the best performance despite it only using three labeled test samples. Our results also show the importance of the base model for the DA method: for the bandgap-LOCO, the neural network based Roost model without fine-tuning is better than all RF-based DA methods. 

In this work, we have only covered the traditional machine learning models and some simple neural network models with DA for realistic materials property prediction. It is known that many state-of-the-art algorithms for the Matbench are based on deep neural networks. Most of the evaluated DA methods do not apply to those complex neural network models except the transfer learning based approaches. We are confident that domain adaptation for material problems has an abundance of potential to significantly improve machine learning performance and are fully confident that it will provide a promising research direction.

\section{Data and Code Availability}

The source code and the non-redundant datasets can be freely accessed at https://github.com/Little-Cheryl/MatDA

\section{Contribution}
Conceptualization, J.H., and R.D.; methodology, J.H., D.L., R.D, N.F.; software, J.H., D.L., R.D., N.F.; writing--original draft preparation, J.H., R.D.; writing--review and editing, R.D., N.F., D.L., J.H.; visualization, J.H., R.D., and N.F.; supervision, R.D.

\bibliographystyle{unsrt}  
\bibliography{references}

\begin{thebibliography}{10}

\bibitem{avery2019predicting}
Patrick Avery, Xiaoyu Wang, Corey Oses, Eric Gossett, Davide~M Proserpio,
  Cormac Toher, Stefano Curtarolo, and Eva Zurek.
\newblock Predicting superhard materials via a machine learning informed
  evolutionary structure search.
\newblock {\em npj Computational Materials}, 5(1):1--11, 2019.

\bibitem{ojih2023screening}
Joshua Ojih, Alejandro Rodriguez, Jianjun Hu, and Ming Hu.
\newblock Screening outstanding mechanical properties and low lattice thermal
  conductivity using global attention graph neural network.
\newblock {\em Energy and AI}, page 100286, 2023.

\bibitem{xin2021active}
Rui Xin, Edirisuriya~MD Siriwardane, Yuqi Song, Yong Zhao, Steph-Yves Louis,
  Alireza Nasiri, and Jianjun Hu.
\newblock Active-learning-based generative design for the discovery of
  wide-band-gap materials.
\newblock {\em The Journal of Physical Chemistry C}, 125(29):16118--16128,
  2021.

\bibitem{chen2020critical}
Chi Chen, Yunxing Zuo, Weike Ye, Xiangguo Li, Zhi Deng, and Shyue~Ping Ong.
\newblock A critical review of machine learning of energy materials.
\newblock {\em Advanced Energy Materials}, 10(8):1903242, 2020.

\bibitem{himanen2020dscribe}
Lauri Himanen, Marc~OJ J{\"a}ger, Eiaki~V Morooka, Filippo~Federici Canova,
  Yashasvi~S Ranawat, David~Z Gao, Patrick Rinke, and Adam~S Foster.
\newblock Dscribe: Library of descriptors for machine learning in materials
  science.
\newblock {\em Computer Physics Communications}, 247:106949, 2020.

\bibitem{jha2019irnet}
Dipendra Jha, Logan Ward, Zijiang Yang, Christopher Wolverton, Ian Foster,
  Wei-keng Liao, Alok Choudhary, and Ankit Agrawal.
\newblock Irnet: A general purpose deep residual regression framework for
  materials discovery.
\newblock In {\em Proceedings of the 25th ACM SIGKDD International Conference
  on Knowledge Discovery \& Data Mining}, pages 2385--2393, 2019.

\bibitem{omee2022scalable}
Sadman~Sadeed Omee, Steph-Yves Louis, Nihang Fu, Lai Wei, Sourin Dey, Rongzhi
  Dong, Qinyang Li, and Jianjun Hu.
\newblock Scalable deeper graph neural networks for high-performance materials
  property prediction.
\newblock {\em Patterns}, 3(5):100491, 2022.

\bibitem{klipfel2023equivariant}
Astrid Klipfel, Zied Bouraoui, Olivier Peltre, Ya{\"e}l Fregier, Najwa Harrati,
  and Adlane Sayede.
\newblock Equivariant message passing neural network for crystal material
  discovery.
\newblock In {\em Proceedings of the AAAI Conference on Artificial
  Intelligence}, volume~37, pages 14304--14311, 2023.

\bibitem{kaba2022equivariant}
Oumar Kaba and Siamak Ravanbakhsh.
\newblock Equivariant networks for crystal structures.
\newblock {\em Advances in Neural Information Processing Systems},
  35:4150--4164, 2022.

\bibitem{choudhary2021atomistic}
Kamal Choudhary and Brian DeCost.
\newblock Atomistic line graph neural network for improved materials property
  predictions.
\newblock {\em npj Computational Materials}, 7(1):185, 2021.

\bibitem{gibson2022data}
Jason Gibson, Ajinkya Hire, and Richard~G Hennig.
\newblock Data-augmentation for graph neural network learning of the relaxed
  energies of unrelaxed structures.
\newblock {\em npj Computational Materials}, 8(1):211, 2022.

\bibitem{chen2021learning}
Chi Chen, Yunxing Zuo, Weike Ye, Xiangguo Li, and Shyue~Ping Ong.
\newblock Learning properties of ordered and disordered materials from
  multi-fidelity data.
\newblock {\em Nature Computational Science}, 1(1):46--53, 2021.

\bibitem{rohr2020benchmarking}
Brian Rohr, Helge~S Stein, Dan Guevarra, Yu~Wang, Joel~A Haber, Muratahan
  Aykol, Santosh~K Suram, and John~M Gregoire.
\newblock Benchmarking the acceleration of materials discovery by sequential
  learning.
\newblock {\em Chemical science}, 11(10):2696--2706, 2020.

\bibitem{wu2019machine}
Stephen Wu, Yukiko Kondo, Masa-aki Kakimoto, Bin Yang, Hironao Yamada, Isao
  Kuwajima, Guillaume Lambard, Kenta Hongo, Yibin Xu, Junichiro Shiomi, et~al.
\newblock Machine-learning-assisted discovery of polymers with high thermal
  conductivity using a molecular design algorithm.
\newblock {\em Npj Computational Materials}, 5(1):66, 2019.

\bibitem{li2023redundancy}
Kangming Li, Daniel Persaud, Kamal Choudhary, Brian DeCost, Michael Greenwood,
  and Jason Hattrick-Simpers.
\newblock On the redundancy in large material datasets: efficient and robust
  learning with less data.
\newblock {\em arXiv preprint arXiv:2304.13076}, 2023.

\bibitem{meredig2018can}
Bryce Meredig, Erin Antono, Carena Church, Maxwell Hutchinson, Julia Ling, Sean
  Paradiso, Ben Blaiszik, Ian Foster, Brenna Gibbons, Jason Hattrick-Simpers,
  et~al.
\newblock Can machine learning identify the next high-temperature
  superconductor? examining extrapolation performance for materials discovery.
\newblock {\em Molecular Systems Design \& Engineering}, 3(5):819--825, 2018.

\bibitem{xiong2020evaluating}
Zheng Xiong, Yuxin Cui, Zhonghao Liu, Yong Zhao, Ming Hu, and Jianjun Hu.
\newblock Evaluating explorative prediction power of machine learning
  algorithms for materials discovery using k-fold forward cross-validation.
\newblock {\em Computational Materials Science}, 171:109203, 2020.

\bibitem{loftis2020lattice}
Christian Loftis, Kunpeng Yuan, Yong Zhao, Ming Hu, and Jianjun Hu.
\newblock Lattice thermal conductivity prediction using symbolic regression and
  machine learning.
\newblock {\em The Journal of Physical Chemistry A}, 125(1):435--450, 2020.

\bibitem{li2023critical}
Kangming Li, Brian DeCost, Kamal Choudhary, Michael Greenwood, and Jason
  Hattrick-Simpers.
\newblock A critical examination of robustness and generalizability of machine
  learning prediction of materials properties.
\newblock {\em npj Computational Materials}, 9(1):55, 2023.

\bibitem{wenzel2022assaying}
Florian Wenzel, Andrea Dittadi, Peter Gehler, Carl-Johann Simon-Gabriel, Max
  Horn, Dominik Zietlow, David Kernert, Chris Russell, Thomas Brox, Bernt
  Schiele, et~al.
\newblock Assaying out-of-distribution generalization in transfer learning.
\newblock {\em Advances in Neural Information Processing Systems},
  35:7181--7198, 2022.

\bibitem{wang2022generalizing}
Jindong Wang, Cuiling Lan, Chang Liu, Yidong Ouyang, Tao Qin, Wang Lu, Yiqiang
  Chen, Wenjun Zeng, and Philip Yu.
\newblock Generalizing to unseen domains: A survey on domain generalization.
\newblock {\em IEEE Transactions on Knowledge and Data Engineering}, 2022.

\bibitem{shen2021towards}
Zheyan Shen, Jiashuo Liu, Yue He, Xingxuan Zhang, Renzhe Xu, Han Yu, and Peng
  Cui.
\newblock Towards out-of-distribution generalization: A survey.
\newblock {\em arXiv preprint arXiv:2108.13624}, 2021.

\bibitem{scholkopf2021toward}
Bernhard Sch{\"o}lkopf, Francesco Locatello, Stefan Bauer, Nan~Rosemary Ke, Nal
  Kalchbrenner, Anirudh Goyal, and Yoshua Bengio.
\newblock Toward causal representation learning.
\newblock {\em Proceedings of the IEEE}, 109(5):612--634, 2021.

\bibitem{wilson2020survey}
Garrett Wilson and Diane~J Cook.
\newblock A survey of unsupervised deep domain adaptation.
\newblock {\em ACM Transactions on Intelligent Systems and Technology (TIST)},
  11(5):1--46, 2020.

\bibitem{zhou2022domain}
Kaiyang Zhou, Ziwei Liu, Yu~Qiao, Tao Xiang, and Chen~Change Loy.
\newblock Domain generalization: A survey.
\newblock {\em IEEE Transactions on Pattern Analysis and Machine Intelligence},
  2022.

\bibitem{farahani2021brief}
Abolfazl Farahani, Sahar Voghoei, Khaled Rasheed, and Hamid~R Arabnia.
\newblock A brief review of domain adaptation.
\newblock {\em Advances in data science and information engineering:
  proceedings from ICDATA 2020 and IKE 2020}, pages 877--894, 2021.

\bibitem{yu2023comprehensive}
Zhiqi Yu, Jingjing Li, Zhekai Du, Lei Zhu, and Heng~Tao Shen.
\newblock A comprehensive survey on source-free domain adaptation.
\newblock {\em arXiv preprint arXiv:2302.11803}, 2023.

\bibitem{de2021adapt}
Antoine de~Mathelin, Fran{\c{c}}ois Deheeger, Guillaume Richard, Mathilde
  Mougeot, and Nicolas Vayatis.
\newblock Adapt: Awesome domain adaptation python toolbox.
\newblock {\em arXiv preprint arXiv:2107.03049}, 2021.

\bibitem{abbasi2020deepcda}
Karim Abbasi, Parvin Razzaghi, Antti Poso, Massoud Amanlou, Jahan~B Ghasemi,
  and Ali Masoudi-Nejad.
\newblock Deepcda: deep cross-domain compound--protein affinity prediction
  through lstm and convolutional neural networks.
\newblock {\em Bioinformatics}, 36(17):4633--4642, 2020.

\bibitem{anastopoulos2021patient}
Ioannis Anastopoulos, Lucas Seninge, Hongxu Ding, and Joshua Stuart.
\newblock Patient informed domain adaptation improves clinical drug response
  prediction.
\newblock {\em bioRxiv}, pages 2021--08, 2021.

\bibitem{jin2020adaptive}
Wengong Jin, Regina Barzilay, and Tommi Jaakkola.
\newblock Adaptive invariance for molecule property prediction.
\newblock {\em arXiv preprint arXiv:2005.03004}, 2020.

\bibitem{wu2022metric}
Fang Wu, Nicolas Courty, Zhang Qiang, Ziqing Li, et~al.
\newblock Metric learning-enhanced optimal transport for biochemical regression
  domain adaptation.
\newblock {\em arXiv preprint arXiv:2202.06208}, 2022.

\bibitem{dunn2020benchmarking}
Alexander Dunn, Qi~Wang, Alex Ganose, Daniel Dopp, and Anubhav Jain.
\newblock Benchmarking materials property prediction methods: the matbench test
  set and automatminer reference algorithm.
\newblock {\em npj Computational Materials}, 6(1):138, 2020.

\bibitem{ward2016general}
Logan Ward, Ankit Agrawal, Alok Choudhary, and Christopher Wolverton.
\newblock A general-purpose machine learning framework for predicting
  properties of inorganic materials.
\newblock {\em npj Computational Materials}, 2(1):1--7, 2016.

\bibitem{goodall2020predicting}
Rhys~EA Goodall and Alpha~A Lee.
\newblock Predicting materials properties without crystal structure: Deep
  representation learning from stoichiometry.
\newblock {\em Nature communications}, 11(1):6280, 2020.

\bibitem{de2021materials}
Pierre-Paul De~Breuck, Geoffroy Hautier, and Gian-Marco Rignanese.
\newblock Materials property prediction for limited datasets enabled by feature
  selection and joint learning with modnet.
\newblock {\em npj Computational Materials}, 7(1):83, 2021.

\bibitem{daume2009frustratingly}
Hal Daum{\'e}~III.
\newblock Frustratingly easy domain adaptation.
\newblock {\em arXiv preprint arXiv:0907.1815}, 2009.

\bibitem{daume2007frustratingly}
Hal Daum{\'e}~III.
\newblock Frustratingly easy domain adaptation. association for computational
  linguistic(acl) s (june): 256--263, 2007.

\bibitem{sun2016return}
Baochen Sun, Jiashi Feng, and Kate Saenko.
\newblock Return of frustratingly easy domain adaptation.
\newblock In {\em Proceedings of the AAAI conference on artificial
  intelligence}, volume~30, 2016.

\bibitem{fernando2013unsupervised}
Basura Fernando, Amaury Habrard, Marc Sebban, and Tinne Tuytelaars.
\newblock Unsupervised visual domain adaptation using subspace alignment.
\newblock In {\em Proceedings of the IEEE international conference on computer
  vision}, pages 2960--2967, 2013.

\bibitem{pan2010domain}
Sinno~Jialin Pan, Ivor~W Tsang, James~T Kwok, and Qiang Yang.
\newblock Domain adaptation via transfer component analysis.
\newblock {\em IEEE transactions on neural networks}, 22(2):199--210, 2010.

\bibitem{uguroglu2011feature}
Selen Uguroglu and Jaime Carbonell.
\newblock Feature selection for transfer learning.
\newblock In {\em Joint European Conference on Machine Learning and Knowledge
  Discovery in Databases}, pages 430--442. Springer, 2011.

\bibitem{sun2016deep}
Baochen Sun and Kate Saenko.
\newblock Deep coral: Correlation alignment for deep domain adaptation.
\newblock In {\em Computer Vision--ECCV 2016 Workshops: Amsterdam, The
  Netherlands, October 8-10 and 15-16, 2016, Proceedings, Part III 14}, pages
  443--450. Springer, 2016.

\bibitem{de2021adversarial}
Antoine de~Mathelin, Guillaume Richard, Fran{\c{c}}ois Deheeger, Mathilde
  Mougeot, and Nicolas Vayatis.
\newblock Adversarial weighting for domain adaptation in regression.
\newblock In {\em 2021 IEEE 33rd International Conference on Tools with
  Artificial Intelligence (ICTAI)}, pages 49--56. IEEE, 2021.

\bibitem{huang2006correcting}
Jiayuan Huang, Arthur Gretton, Karsten Borgwardt, Bernhard Sch{\"o}lkopf, and
  Alex Smola.
\newblock Correcting sample selection bias by unlabeled data.
\newblock {\em Advances in neural information processing systems}, 19, 2006.

\bibitem{yamada2013relative}
Makoto Yamada, Taiji Suzuki, Takafumi Kanamori, Hirotaka Hachiya, and Masashi
  Sugiyama.
\newblock Relative density-ratio estimation for robust distribution comparison.
\newblock {\em Neural computation}, 25(5):1324--1370, 2013.

\bibitem{kanamori2009least}
Takafumi Kanamori, Shohei Hido, and Masashi Sugiyama.
\newblock A least-squares approach to direct importance estimation.
\newblock {\em The Journal of Machine Learning Research}, 10:1391--1445, 2009.

\bibitem{loog2012nearest}
Marco Loog.
\newblock Nearest neighbor-based importance weighting.
\newblock In {\em 2012 IEEE International Workshop on Machine Learning for
  Signal Processing}, pages 1--6. IEEE, 2012.

\bibitem{wu2004improving}
Pengcheng Wu and Thomas~G Dietterich.
\newblock Improving svm accuracy by training on auxiliary data sources.
\newblock In {\em Proceedings of the twenty-first international conference on
  Machine learning}, page 110, 2004.

\bibitem{chelba2006adaptation}
Ciprian Chelba and Alex Acero.
\newblock Adaptation of maximum entropy capitalizer: Little data can help a
  lot.
\newblock {\em Computer Speech \& Language}, 20(4):382--399, 2006.

\bibitem{segev2016learn}
Noam Segev, Maayan Harel, Shie Mannor, Koby Crammer, and Ran El-Yaniv.
\newblock Learn on source, refine on target: A model transfer learning
  framework with random forests.
\newblock {\em IEEE transactions on pattern analysis and machine intelligence},
  39(9):1811--1824, 2016.

\bibitem{pardoe2010boosting}
David Pardoe and Peter Stone.
\newblock Boosting for regression transfer.
\newblock In {\em Proceedings of the 27th International Conference on
  International Conference on Machine Learning}, pages 863--870, 2010.

\bibitem{dai2007boosting}
Wenyuan Dai, Qiang Yang, Gui-Rong Xue, and Yong Yu.
\newblock Boosting for transfer learning.
\newblock In {\em Proceedings of the 24th international conference on Machine
  learning}, pages 193--200, 2007.

\bibitem{varivoda2023materials}
Daniel Varivoda, Rongzhi Dong, Sadman~Sadeed Omee, and Jianjun Hu.
\newblock Materials property prediction with uncertainty quantification: A
  benchmark study.
\newblock {\em Applied Physics Reviews}, 10(2), 2023.

\end{thebibliography}

\end{document}